\documentclass[12pt]{article}

\usepackage{amsfonts}
\usepackage{amssymb}
\usepackage{amsmath}
\usepackage{amsthm}
\usepackage{eurosym}
\usepackage{upgreek}
\usepackage{tikz}
\usepackage{graphicx}
\usepackage{multirow}
\usepackage{subfigure}
\usepackage{pgfplots}
\usepackage{units}
\RequirePackage[colorlinks,citecolor=blue,linkcolor=red,urlcolor=blue]{hyperref}
\RequirePackage{natbib}

\usetikzlibrary{shapes,snakes}

\theoremstyle{theorem}
\newtheorem{theorem}{Theorem}

\newtheorem{corollary}{Corollary}

\newtheorem{lemma}{Lemma}
\newtheorem{exam}{Example}
\newenvironment{example}{\begin{exam} \rm }{\hfill  $\triangleleft$ \end{exam}}\newtheorem{des}{Design}
\newenvironment{design}{\begin{des} \rm }{\hfill  $\triangleleft$ \end{des}}
\newtheorem*{exam1}{Example 1 continued}

\newtheorem*{exam2}{Example 2 continued}

\newtheorem*{exam3}{Example 3 continued}

\newtheorem{assum}{Assumption}

\newtheorem{rema}{Remark}

\newtheorem{defin}{Definition}
\newenvironment{definition}{\begin{defin} \rm }{\hfill  \end{defin}}
\newtheorem{mechan}{Mechanism}

\pgfplotsset{my style/.append style={axis x line=middle, axis y line=
           middle}}

\DeclareMathOperator{\inter}{int}

\DeclareFontFamily{U}{mathx}{\hyphenchar\font45}
\DeclareFontShape{U}{mathx}{m}{n}{
      <5> <6> <7> <8> <9> <10>
      <10.95> <12> <14.4> <17.28> <20.74> <24.88>
      mathx10
      }{}
\DeclareSymbolFont{mathx}{U}{mathx}{m}{n}
\DeclareMathSymbol{\bigtimes}{1}{mathx}{"91}

\makeatletter

\renewcommand*{\@seccntformat}[1]{%
  \csname the#1\endcsname.\quad}
\makeatother

\begin{document}

\title{Belief identification with state-dependent utilities\footnote{An earlier version of this paper circulated under the title ``Identification of misreported beliefs". I am greatly indebted to Jean Baccelli, Aurelien Baillon, Georgios Gerasimou, Itzhak Gilboa, Jeanne Hagenbach, Edi Karni, Hannes Leitgeb, Jay Lu, Andy Mackenzie, Sasha Vostroknutov, Peter Wakker and Roberto Weber for their instrumental comments at various stages of this project. I would also like to thank the audiences at FUR (Ghent), LOFT (Groningen), Workshop on strategic communication, bounded rationality and complexity (Cergy), Epicenter Workshop (Maastricht), RCEA (online) and internal seminars in Maastricht University for useful input.}}


\author{
\textsc{Elias Tsakas}\footnote{Department of Microeconomics and Public Economics, Maastricht University, P.O. Box 616, 6200 MD, Maastricht, The Netherlands; Homepage: \url{www.elias-tsakas.com}; E-mail: \href{mailto:e.tsakas@maastrichtuniversity.nl}{\texttt{e.tsakas@maastrichtuniversity.nl}}}\\ 
\small{\textit{Maastricht University}}}

\date{December, 2022}

\maketitle

\begin{abstract}

\noindent It is well known that individual beliefs cannot be identified using traditional choice data, unless we impose the practically restrictive and conceptually awkward assumption that utilities are state-independent. In this paper, we propose a novel methodology that solves this long-standing identification problem in a simple way, using a variant of the strategy method. Our method relies on the concept of a suitable proxy. The crucial property is that the agent does not have any stakes in the proxy conditional on the realization of the original state space. Then, instead of trying to identify directly the agent's beliefs about the state space, we elicit her conditional beliefs about the proxy given each state realization. The latter can be easily done with existing elicitation tools and without worrying about the identification problem. It turns out that this is enough to uniquely identify the agent's beliefs. We present different classes of proxies that one can reasonably use, and we show that it is almost always possible to find one which is easy to implement. Such flexibility makes our method, not only theoretically-sound, but also empirically appealing.  We also show how our new method allows us to provide a novel well-founded definition of a utility function over states. Last but not least, it also allows us to cleanly identify motivated beliefs, freed from confounding distortions caused by the identification problem.

\vspace{0.5\baselineskip}



\end{abstract}

\section{Introduction}

\subsection{Problem statement} 

Identifying subjective beliefs is a central problem in economics that dates back to the early seminal contributions of \cite{Ramsey1931}, \cite{DeFinetti1937} and \cite{Savage1954}. While this literature originally focused mostly on providing foundations for the notion of subjective probability, more recently economists have also recognized the practical importance of the question \citep{Manski2004}. This renewed interest comes from the fact that beliefs are nowadays widely used for a broad range of purposes, e.g., to make out-of-sample predictions; to obtain, compare, and aggregate forecasts for events of interest; to study irregularities in how people process information, etc. Therefore, being able to measure actual beliefs accurately is of outmost importance for applied research and policy.\footnote{Throughout the paper we assume that an actual belief exists, and it is therefore interpreted as an unobservable primitive. Still, we show that our entire analysis can be easily adapted to settings where the belief is interpreted as a parameter that only acquires meaning within a SEU model (Section \ref{S:Definition of subjective probability}).}


The traditional economics methodology relies on identifying beliefs through observed betting behavior. Specifically, an agent's choices among acts (i.e., state-contingent payments) are supposed to reveal her beliefs over the state space \citep{Savage1954, AnscombeAumann1963, Wakker1989}. However, there is a foundational caveat, known as the \textit{identification problem}, which undermines this methodology \citep[e.g.,][]{Dreze1961,Dreze1987,Fishburn1973, KarniSchmeidlerVind1983}: 
\begin{quote}
\textit{The agent's beliefs can be identified through traditional choices over acts, only if we exogenously assume her utility function to be state-independent.}
\end{quote}
Let us illustrate the problem by means of an example.\footnote{This is a modification of the standard example, which was originally used in a letter correspondence between Bob Aumann and Leonard Savage \citep{Dreze1987}, and has been used widely in this literature since then. We will use this story as our running example throughout the paper. For a formal presentation, see Example \ref{EX:insurance problem}.} Suppose that a man suffers from Guillain-Barr\'{e} syndrom, a serious neurological condition that has left him completely paralyzed, and it is unclear whether he will recover from the disease within the next year. His wife attaches subjective probability $10\%$ to her husband recovering, which is something known only to her. Suppose that we want to identify this probability. Since she loves her husband very much, she values extra money more in a world where he has recovered and she can spend it with him, compared to a world where he is paralyzed. Hence, we cannot assume her utility over monetary payoffs to be state-independent. As a result, the relationship between her beliefs and her willingness to accept a bet is confounded by her relative utility for money between the two states. And since the latter is no longer normalized to 1 ---as it would have been under the state-independence assumption--- we cannot identify beliefs from her betting behavior. 

This poses a serious problem. For starters, in many economic applications where we typically elicit beliefs, utility functions are state-dependent, either due to intrinsic preferences over states or due to unobservable state-dependent side payoffs, e.g., in insurance problems \citep{Arrow1974, CookGraham1977, DrezeRustichini2004, Karni2008b}, legal judgments \citep{Andreoni1991,FeddersenPesendorfer1998, Tsakas2017}, real estate decisions \citep{CaseShiller2012}, medical decisions \citep{PaukerKassirer1975,PaukerKassirer1980,Lu2019}, to mention a few. Moreover, widely-studied phenomena like political polarization are typically explained by psychological theories that crucially rely on assuming utilities to be state-dependent, e.g., motivated reasoning  \citep{Kunda1990, Benabou2015} or ingroup favoritism \citep{Everett2015}. As a result, in all these settings we cannot be confident that we measure the actual beliefs, and therefore our conclusions are in principle confounded by this measurement problem. 

Before moving forward, it is important to point out that according to an early view among theorists, the identification problem is not really a problem. Their argument is that state-independent utilities are simply a normalization that conveniently allows us to decompose tastes from probabilities in our SEU representation, and these probabilities may very well differ from the actual beliefs. But we should not really care about this discrepancy, as it will not affect anyway our ability to predict choices among acts. In this sense, according to this view, it is questioned why we would want to identify the actual beliefs in the first place \citep[for an overview, see][and references therein]{Karni2014}. On the flip side, our counterargument is that actual beliefs are often used for a wide range of purposes that go beyond predicting choices among acts. For instance, an investor wants to know the actual beliefs of a real estate expert about future house prices, not because she wants to predict whether the expert will buy a house for himself, but rather in order to use this information for her own downstream investment decision. Likewise, an experimental economist wants to know the actual beliefs of Democrats and Republicans about the winner of the next presidential election, not because she cares about how these subjects would be betting in a prediction market, but rather in order to test the hypothesis that beliefs diverge due to motivated reasoning. And it is exactly in such settings where the identification problem has serious consequences \citep[see][and references therein]{Baccelli2017}.

Historically, the identification problem was already noticed during the very early days of decision theory: as Leonard Savage admits in his well-known letter correspondence with Bob Aumann, ``the problem is serious, but I am willing to live with it until something better comes along" \citep{Dreze1987}. What is hidden behind these simple words is a fundamental tradeoff. Namely, in order to identify an agent's beliefs we must either accept the conceptually awkward assumption of state-independent utilities (the \textit{practically-oriented approach}), or we must go well beyond traditional choice data (the \textit{theoretically-sound approach}). The former is usually adopted by experimental economists, who ---in the name of simplicity--- disregard the problem. And although at the outset this may sound like a simplistic approach, it is actually quite pragmatic. Indeed, even though decision-theorists have proposed various solutions that use non-traditional choice data \citep[e.g.,][among others]{Dreze1961,Dreze1987, Fishburn1973, KarniSchmeidlerVind1983, Karni1992, Lu2019}, none of them is unanimously accepted as the standard one. This suggests that the identification problem is both very hard and still open.

\subsection{Our contribution}

The main reason why none of the existing attempts to overcome the identification problem has been broadly adopted as ``the standard solution" is that non-traditional choice data are in general complex and cumbersome. And this is exactly where the core of our contribution lies. Namely, we propose a novel solution that sticks to the standard methodology of using traditional choice data, albeit over an extended state space. In this sense, our solution will turn out to be not only theoretically sound but also tractable.

The key is the way we extend the state space. We do this by taking a product space, where one dimension is the original state space and the second dimension is what we call a \textit{suitable proxy}. Then, instead of observing the agent's betting behavior over the original state space, we will look at her betting behavior over the proxy conditional on realizations of the original state space. In practice, this is simply an application of the strategy method, in the sense that the agent will be asked to make a choice conditional on each state realization. Remarkably, this will turn out to suffice for identifying her beliefs about the original state space, which is what we have been interested in all along. Let us illustrate the method in the context of the wife's problem that we presented in the previous section. 

Suppose that there is a promising new experimental drug in the market which is strongly believed to expedite the recovery time from Guillain-Barr\'{e}. The husband is eligible to participate in the last phase of the clinical trial. The proxy describes the two possible contingencies regarding his participation, i.e., he will receive either the drug $(t_1)$ or a placebo $(t_2)$. The wife is informed that his chances of being placed in the treatment group are 50\%, but she is not told in which group he is actually placed in the end. Of course, if she were to learn that he received the drug, her belief would change to some $\nu$; on the other hand, if she learned that he received the placebo, her belief would remain unchanged at $\mu$, which is the one we have been trying to identify in the first place.\footnote{Here, we implicitly assume that the husband does not even know that he is participating in the clinical trial, and therefore his health cannot be affected by psychological factors, such as the well-known placebo effect.} For a graphical illustration, see Figure \ref{FIG:proxies graphical representation} below.

\begin{figure}[h!]
\begin{center}
\begin{tikzpicture}[scale=1.2]
\draw[->] (0,1) -- (1.8,0);
\draw[->] (0,1) -- (1.8,2);
\draw (2.3,0) node[left] {\footnotesize{$t_2$}};
\draw (2.3,2) node[left] {\footnotesize{$t_1$}};
\draw[->] (2.3,0) -- (4.3,0);
\draw[->] (2.3,0) -- (4.3,1.9);
\draw[->] (2.3,2) -- (4.3,0.1);
\draw[->] (2.3,2) -- (4.3,2);
\draw (4.3,0) node[right] {\footnotesize{$s_2:$ the husband remains paralyzed}};
\draw (4.3,2) node[right] {\footnotesize{$s_1:$ the husband recovers}};
\draw (0.7,1.8) node {\footnotesize{$50\%$}};
\draw (0.7,0.2) node {\footnotesize{$50\%$}};
\draw (3,2) node[above] {\footnotesize{$\nu_1$}};
\draw (3,0) node[below] {\footnotesize{$\mu_2$}};
\draw (2.6,0.6) node {\footnotesize{$\mu_1$}};
\draw (2.6,1.4) node {\footnotesize{$\nu_2$}};
\end{tikzpicture}
\end{center}
\vspace{-1\baselineskip}
\caption{\footnotesize{The two contingencies that the proxy describes are ``the husband takes the experimental drug $(t_1)$" and ``the husband takes the placebo $(t_2)$". The prior probability of the husband (randomly) placed in the control group is 50\%. Conditional on the placebo being taken, the wife's belief remains at $\mu$. Conditional on the drug being taken, her belief changes to $\nu$.}}
\label{FIG:proxies graphical representation}
\end{figure}
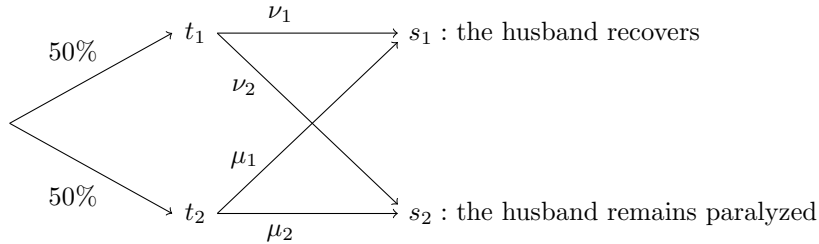

The crucial feature is that the wife \textit{does not have any stakes in the realization of the proxy conditional on her husband's health status}. For instance, conditional on him recovering (resp., remaining paralyzed), she will not care about whether he has taken the drug or the placebo, and therefore her utility from money will not depend on the realization of the proxy.\footnote{The implicit assumption that we make for the sake of our simple example is that the drug is known not have any side-effects that would make the wife care about whether her husband has taken it or not conditional on eventual health condition.} Hence, we can identify her conditional beliefs of him having taken the placebo given that her husband has recovered, as well as given that he has remained paralyzed. In both cases, we can do this, using standard elicitation tasks, \textit{without worrying about the identification problem}. 

As it turns out, this information is enough for identifying the wife's actual belief about her husband's recovery. The idea is very simple. The wife's belief $\pi_T$ of her husband taking the placebo has been publicly announced to her, and therefore we know it. Her beliefs $\pi_T(\cdot|s_1)$ and $\pi_T(\cdot|s_2)$ of him taking the placebo conditional on his health status have just been identified, as explained in the previous paragraph. Thus, if we interpret Figure \ref{FIG:proxies graphical representation} as an information structure, we know the prior and posterior beliefs. Hence, we can easily identify the likelihoods $\nu$ and $\mu$. But the latter is exactly the actual belief that we have been trying to identify all along, i.e., her belief of him recovering given that he has taken the placebo is the same as her belief of him recovering without having introduced the proxy in the first place.

Our main result (Theorem \ref{T:Main Theorem}) generalizes this idea, showing that \textit{the previous method identifies the actual belief over any finite state space for a broad range of proxies}. The latter is particularly important, as it addresses the issue of applicability of our method. Namely, one can naturally ask how easy it is to find a suitable proxy. For instance, in the context of our earlier example, what can we do if there is no clinical trial taking place at this moment? As it turns out, it is almost always possible to find a suitable proxy that satisfies the conditions of our theorem, and a fortiori allows us to identify the beliefs. Let us go back to our running example.

Suppose that the wife is told that some doctor predicted that her husband will recover. But she is also told that there is probability 50\% that this doctor is a charlatan whose diagnoses are pure noise. In this case, the proxy describes the two possible contingencies regarding the expertise of the doctor, or equivalently regarding the informativeness of the diagnosis, i.e., the doctor is either an expert $(t_1)$ or a charlatan $(t_2)$. Of course, if the wife learned that the diagnosis has come from an expert, she would somehow update her belief; whereas, if she learned that it has come from a charlatan, she would keep her prior belief. Crucially, the wife will not have any stakes in the expertise of the doctor conditional on her husband's state of health having been realized. Then, using the same argument as in the example with the clinical trial, we can infer her conditional belief about his recovery given that the doctor is a charlatan, which is the same as her belief about his recovery in the absence of any diagnosis. This is because the information structure in Figure \ref{FIG:proxies graphical representation} can also be used in exactly the same way in this example. 

What the two aforementioned examples represent are two big classes of suitable proxies. The first one includes what we call \textit{influential actions}. These are stochastic actions controlled by the analyst which directly interfere with the realization of the original state space. The second class includes what we call \textit{evidence with stochastic reliability}. These are signals which are informative with some exogenously given probability. Obviously, while the former is often difficult and perhaps too costly to implement, the latter is almost always possible. Furthermore, there is third class of suitable proxies, which we call \textit{random sampling}. Such proxies are useful when we want to identify the agent's beliefs about the expected value of some variable, e.g., what does the CEO of the pharmaceutical company believe regarding the expected number of patients that the drug will cure, or what does an advisor of a presidential candidate believe about the expected percentage of votes that the candidate will receive. In such cases, a proxy in the form of random sampling relies on the agent knowing the distribution of some other relevant demographic characteristic within the population, e.g., distribution of age within patients, or the distribution of gender within voters. We further elaborate on the different classes of proxies in Section \ref{S:extending the state space}, and we provide several examples in Section \ref{S:finding suitable proxies}. 

Generally speaking, the abundance of potentially suitable proxies provides us with a lot of flexibility, and it is exactly this flexibility that makes our approach appealing. To see why this is the case, first observe that methodologically our approach is not that different from the standard approach which is commonly used in practice, viz., both observe betting behavior over some state space, and both need to make some exogenous non-testable assumption in order to infer the agent's beliefs from this behavior. In particular, the standard approach assumes that the agent does not have any stakes in the original state space, whereas we assume that the agent does not have any stakes in the proxy conditional on the original state space. The crucial difference is that the standard approach imposes this assumption ex ante, before having specified the underlying state space, and this is why in some cases it is difficult to justify it. In our approach, on the other hand, we impose the assumption ex post, viz., \textit{we only pick the proxy once we already know the original state space}. As a result, it becomes much easier to choose a suitable proxy for which we are quite confident that the state-independence assumption is conditionally justified.\footnote{A good analogy to this comparison comes from econometrics. The assumption of explanatory variables being independent of the error term is a default assumption that we often impose ex ante, before having decided on the final specification of our model. On the other hand, instrumental variables are specifically chosen ex post to deal with the endogeneity problem once the model has been already specified.} 
 
Then, once we have identified the actual belief, we can also uniquely identify the agent's ``actual" (state-dependent) utility function, using only traditional choice data (Theorem \ref{THM:actual utility function}). The interesting consequence of this result is that it allows us to obtain a preference ordering over the state space. The latter allows us to formally attach meaning to statements like ``the wife prefers her husband recovering over remaining paralyzed", which are commonly used in applied literature, albeit only in an informal way. 

Of course, being able to uniquely identify an ``actual" utility function once a belief has been pinned down is not something unique to our model, and it is shared by all papers on state-dependent SEU. Nevertheless, to the best of our knowledge, none of these papers uses the duality between beliefs and utilities to define actual preferences over states. The reason is twofold. First, as we have already mentioned, some of the existing literature rejects the idea that there is an actual belief, and a fortiori an ``actual" utility function. As a result, even talking about preferences over states is meaningless. Second, even according to authors who share our view of actual beliefs existing, if a state-independent SEU representation exists, the actual belief will be the one given by this representation. This means that the ``actual" utility function is state-independent, and thus the agent is automatically assumed to be indifferent across all states. As a result, the common understanding within this literature is that preferences over states become meaningful only when there is no state-independent representation. In our paper, on the other hand, there is no such dichotomy. In particular, preferences over states are defined regardless of whether a state-independent SEU representation exists or not. A good example is the wife's problem: despite the fact that a state-independent SEU representation exists, our approach will tell us that the wife prefers her husband to recover than to remain paralyzed. We further elaborate on this point in Section \ref{S:ranking the states}.

One obvious applied setting where agents have stakes in the state realization is the entire literature on motivated beliefs. As such, this is a prime suspect for an environment where the identification problem kicks in. This does not mean that the phenomenon of self-serving beliefs is non-existent, but rather that it may very well be confounded by measurement errors due to the identification problem. Fortunately, our results allow us to disentangle belief distortions due to motivated reasoning from distortions due to measurement errors. And notably, this can be done without needing to resort to some exotic elicitation task. We further discuss this point in Section \ref{S:motivated beliefs}.

The final point that we would like to make is that although throughout our paper we assume a rational agent who updates beliefs in a Bayesian manner, our approach can be easily adjusted to allow for non-Bayesian updating (Section \ref{S:calibrated beliefs}). This would require additional out-of-sample data to calibrate updating parameters, which we can then use to identify the agent's belief using a slight modification of our main theorem. Methodologically, this approach is similar to the one in \cite{Offerman2009}, in the sense that we keep using our original simple mechanism and then we debias the elicited beliefs ex post using out of sample data.

\subsection{Literature review}

There is a lot of existing work dealing with the identification problem. This literature is roughly split in two streams, one that focuses on providing tools which can be potentially used in practice for belief elicitation purposes, and another one which is seen as part of axiomatic decision theory and focuses mostly on the conditions under which belief identification can be achieved theoretically. For a more complete account of this literature, we refer to the reviews of \cite{DrezeRustichini2004}, \cite{Karni2008, Karni2014} and \cite{Baccelli2017}. 

Starting with the first stream, the only papers that introduce mechanisms for eliciting beliefs under state-dependent preferences are \cite{Karni1999} and \cite{JaffrayKarni1999}, with the latter proposing two different mechanisms. In particular, \cite{Karni1999} and the first mechanism of \cite{JaffrayKarni1999} rely on the assumption that state utilities are bounded, and they approximate the actual beliefs in the limit as monetary incentives grow arbitrarily large.\footnote{For an extensive discussion on the boundedness of the utility function, see \cite{Wakker1993}.} This is a rather uncomfortable convention, as the elicitation task will rely on a very large dataset. Moreover, we will either need to incur an extremely high cost, or to use hypothetical data. These problems are recognized by the authors of the two aforementioned papers, who point out that in those early days of the literature there was no other option \citep[e.g.,][p.485]{Karni1999}. The second mechanism in \cite{JaffrayKarni1999} assumes that state-dependence enters the picture in terms of unobserved state-dependent payments. So, first, it proceeds to elicit these payments, and once these are known, it goes on to elicit beliefs using standard techniques. Of course, this is a rather restrictive setting: in most interesting applications, preferences over states are intrinsic. Besides, eliciting the state-dependent payments is quite demanding in terms of the amount of data that is needed.

Turning to the second stream, the various attempts within axiomatic decision theory differ in terms of the non-traditional choice domain they consider. Within this stream, there are three main methodological approaches, all of which rely on the agent's beliefs being somehow revised at some instance. 

The first such method, which was mainly followed in the early days, was based on exogenously manipulating context and a fortiori the agent's beliefs. For instance, \cite{Fishburn1973} allows for comparison between acts conditional on different events. \cite{KarniSchmeidlerVind1983} and \cite{KarniSchmeidler2016} introduce hypothetical preferences over acts, conditional on exogenously given probabilities over the states. \cite{Kadane1990} allow the agent to compare lotteries at different states. \cite{Karni1992, Karni1993} allows the analyst to observe preferences conditional on different events. The main difference of this work in comparison to ours is that they all assume the analyst to observe the agent's choices under different hypothetical contingenciesl. This feature makes the implementation of these methods practically cumbersome.

The second method relies on the idea that the agent herself can affect the state realization, and is called the \textit{moral hazard approach}. In fact, it originally appeared in the early sixties \citep{Dreze1961}, before resurfacing again a couple of decades later \citep{Dreze1987, DrezeRustichini1999} and recently receiving attention again \citep{Baccelli2021}. This literature relates to our work on a high level. Of course the difference is that according to this approach the analyst needs to rely on the agent influencing the state realization, whereas in our case the analyst has full control of the situation. The second major difference is that the moral hazard methodology applies only to settings where the state has not been realized yet, while our method does not pose any such restriction, i.e., our theory also applies to factual beliefs. 

The third and final method is more recent, and relies on the agent updating her beliefs using information that the analyst provides \citep{Lu2019}. This is admittedly a very promising method, similar in spirit to our proxies in the form of evidence with stochastic reliability. In particular, similarly to our work, it can be potentially applied both in cases where the state has been already realized, as well as in cases where it has not.  It does not impose restrictive assumptions at the outset. The only potential drawback that we see, is that it requires stochastic choices (under different information structures), implying that one would need a rather large number of observations. Nevertheless, we still see it as a very promising alternative methodology.

Somewhere in between the second and the third method, one can place a sequence of papers that rely on the idea that the agent can influence the state realization and choose an act, conditional on different signals \citep{Karni2011a,Karni2011b,Karni2014}. The common element across these papers of Karni, the aforementioned paper of Lu, and our work is that all can be thought to rely on traditional choice data over an extended state space. Nevertheless, the way this general idea is implemented is very different.

\subsection{Structure of the paper}

Section \ref{S:identification problem} presents the background and formally introduces the identification problem. Section \ref{S:solution by proxy} presents our solution by proxy and formally states our main results. In Section \ref{S:applications} we present some applications, and in particular our definition of preferences over states and our contribution to identifying motivated beliefs. Section \ref{S:implementation} discusses some issues that pertain to the practical implementation of our proposed methodology. All proofs are relegated to Section \ref{S:proofs} in the Appendix. Section \ref{S:decision-theoretic foundations} in the Appendix formally presents decision-theoretic foundations and relates our work to the existing axiomatic literature.


\section{The identification problem}\label{S:identification problem}

\subsection{The problem statement}\label{S:the problem statement}

Consider a finite state space $S=\{s_1,\dots,s_K\}$. A (female) agent has a (full-support) belief $\mu\in\Delta(S)$. Throughout the paper, we will treat her belief as an unobservable primitive, i.e., this is an actual belief that quantifies her uncertainty over the state space.\footnote{In Section \ref{S:Definition of subjective probability} we will revisit our model under the alternative interpretation of beliefs, which is often found in axiomatic decision theory, according to which $\mu$ is simply an endogenous parameter that obtains meaning only within the context of an expected utility model.} Our goal is to identify this belief through observed choices.

The choice domain which is traditionally used for this purpose is the set of acts $\mathcal{F}_S$. An act is a function $f:S\rightarrow Q$ that maps each state $s$ to a consequence $f_s:=f(s)$ in a convex set $Q$. The agent is assumed to have preferences $\succeq$ over $\mathcal{F}_S$ that admit a \textit{state-dependent Subjective Expected Utility (abbrev., SEU)} representation.\footnote{In Section \ref{S:foundations state-dependent SEU} we discuss in detail the decision-theoretic foundations of state-dependent SEU.} That is, there exists some state-dependent utility function $u:Q\rightarrow\mathbb{R}^S$, such that the SEU function 
\begin{equation}\label{EQ:SDSEU}
\mathbb{E}_{\mu}(u(f)):=\sum_{s\in S} \mu(s) u_s(f_s)
\end{equation}
preserves the preference order, i.e., formally for any two acts $f,g\in\mathcal{F}_S$,
\begin{equation}\label{EQ:state-dependent SEU representation}
f\succeq g \ \Leftrightarrow \ \mathbb{E}_\mu(u(f))\geq \mathbb{E}_\mu(u(g)).\end{equation}

The general idea behind a SEU representation is to obtain a model of choice that disentangles beliefs from utilities. Is this enough for identifying the agent's actual beliefs? Unfortunately, not! The reason is that the pair $(u,\mu)$ is not the only SEU representation. Namely, take any full-support belief $\tilde{\mu}\in\Delta(S)$ and define the rescaled utility functions
\begin{equation}\label{EQ:utility rescaling}
\tilde{u}_s=\frac{\mu(s)}{\tilde{\mu}(s)}u_s.
\end{equation}
Then, the pair $(\tilde{u},\tilde{\mu})$ will constitute an alternative SEU representation of the same preferences. This is because, for every act $f\in\mathcal{F}_S$, we will have 
\begin{equation}
\mathbb{E}_{\tilde{\mu}}(\tilde{u}(f))=\mathbb{E}_\mu(u(f)). 
\end{equation}
So, even in the extreme scenario where we manage to observe the complete preference relation over acts, we will not be able to tell if the actual belief is $\mu$ or $\tilde{\mu}$. That is, formally, beliefs cannot be identified from traditional choice data (i.e., from observed choices over $\mathcal{F}_S$).  This is known as the  \textit{identification problem} of SEU.

\subsection{State-independent representations}\label{S:problem state-independent representations}

The early solution to the identification problem was to exogenously assume a state-independent utility function \citep{Savage1954, AnscombeAumann1963,Wakker1989}. Essentially, this approach boils down to appropriately normalizing every state utility function to $u_s=\bar{u}$ in a way such that the tuple $(\bar{u},\bar{\mu})$ is a SEU representation, and then assuming that $\bar{\mu}$ is the actual belief. Admittedly, this seems uncontroversial in settings where the agent does not have a stake in the state realization. However, it is often conceptually difficult to defend it in cases where the agent cares about the state realization, as illustrated in the following example which we have already described in the introduction. 



\begin{example}\label{EX:insurance problem}\textsc{(The wife's problem).}
A man suffers from Guillain-Barr\'{e} syndrom, and it is unclear whether he will recover from the disease within the next year (state $s_1$) or he will remain paralyzed (state $s_2$). His risk-neutral wife attaches subjective probability $\mu(s_1)=10\%$ to her husband recovering, which is something known only to her. Nevertheless, when their insurance company offers her an insurance with payout of \$100k in case he remains paralyzed, the maximum premium she is willing to pay is just \$10k. If we assumed a state-independent utility function, we would conclude that the subjective probability she attaches to her husband recovering would be equal to $\bar{\mu}(s_1)=90\%$. This is because the model wrongly treats the two states symmetrically. That is, it assumes the same linear utility function, $\bar{u}_1(q)=\bar{u}_2(q)=\gamma q$ with $\gamma>0$ at both states. Now, consider instead a SEU model $(\tilde{u},\tilde{\mu})$ with state utility functions $\tilde{u}_1(q)=\gamma_1 q$ and $\tilde{u}_2(q)=\gamma_2 q$ with $\gamma_1>\gamma_2>0$. The interpretation is clear: the wife values each additional dollar more when her husband is healthy than when he is lying paralyzed in bed. Then, her willingness to buy the insurance implies that the model parameters must satisfy the following condition:
\begin{equation}\label{EQ:Example husband}
\frac{\tilde{\mu}(s_2)}{\tilde{\mu}(s_1)}=\frac{\gamma_1}{9\gamma_2}.
\end{equation}
In other words, her actual belief $\mu(s_1)=10\%$, would have been identified correctly if we had set $\gamma_1=81\gamma_2$. Nevertheless, the choice of $\gamma_1$ and $\gamma_2$ is arbitrary, viz., there is no reason ex ante to pick this specific model among all SEU representations. Hence, our belief identification problem is essentially a model specification problem.
\end{example}

Despite the difficulty to justify it, one common argument in favor of adopting a state-independent SEU representation is that we should not care about the actual belief anyway: all SEU representations predict exactly the same choices among acts, which is what matters after all, e.g., in the previous example, every SEU representation would predict that the agent turns down the insurance plan if and only if the premium in higher than \$10k. Our counterargument here is that belief identification is often important for purposes that go well beyond predicting choices over acts. In Example \ref{EX:insurance problem}, suppose that the husband's current employer is willing to extend his current contract only if his recovery probability is larger than 80\%. To this end the employer wants to use the wife's actual belief as a good approximation of her husband recovery probability. So, from the employer's point of view it is crucial to select a SEU representation that involves her actual beliefs. And from our previous analysis it is clear that the state-independent SEU representation would constitute a misspecified model that would lead the employer to make a bad decision.

In fact, to make things even more complicated, oftentimes a state-independent SEU representation does not even exist in the first place. This is for instance the case when the (ordinal) preferences over consequences differ across states, or when there are unobservable state-contingent side payments to the agent.


Concluding, unless it is obvious that the agent has no stakes in the state space, it is difficult to defend the assumption of state-independent utilities, and therefore it remains unclear which SEU representation one should use in order to identify the agent's beliefs. 

\subsection{Belief elicitation mechanisms}\label{S:belief elicitation mechanisms}

The aforementioned identification problem explains why traditional elicitation mechanisms are in general uninformative without the assumption of state-independent utilities. A belief elicitation mechanism incentivizes the agent by rewarding accurate reports and punishing inaccurate ones. Formally, it can be written as a function $\Delta(S)\mapsto\mathcal{F}_S$, that takes as input the agent's reported belief $r\in\Delta(S)$, and returns as output a consequence $r(s)\in Q$ at each state $s\in S$. Existing mechanisms in the literature differ in how the incentives are framed, and whether the consequences are sure monetary payoffs or lotteries. For instance, scoring rules are direct mechanisms that present these incentives explicitly: in their simple version, they pay in monetary amounts \citep{Brier1950,Good1952,Savage1971}, whereas in their binarized version they pay in probabilities to win a fixed prize \citep{HossainOkui2013}. On the other hand, the Karni mechanisms \citep{Karni2009,Tsakas2019} and the method of matching probabilities \citep{DucharmeDonnell1973, KadaneWinkler1988, Baillon2018} propose different ways to associate each report with a compound act, which can in turn be reduced into a simple act. For an overview, see the review articles of \cite{SchlagWeele2013} and \cite{SchotterTrevino2014}, and references therein.

An elicitation mechanism is said to be incentive-compatible if it guarantees truthfull revelation of the actual belief. Formally, for any actual belief $\mu\in\Delta(S)$ and any report $r\neq \mu$, the act associated with reporting truthfully is strictly preferred to the act associated with any other report, i.e., $\mu\succ r$. Unfortunately, the identification problem implies that no such mechanism exists. To see this, take two agents with the same preferences $\succeq$ and different beliefs, $\mu$ and $\tilde{\mu}$ respectively. Then obviously, for any elicitation mechanism, we cannot have both $\mu\succ\tilde{\mu}$ and $\tilde{\mu}\succ\mu$, meaning that at least one of the two agents will have an incentive to misreport, and therefore no mechanism is not incentive-compatible. Of course, this is not surprising given that each mechanism is actually a menu of acts. And as we have already discussed, observing choices among acts does not suffice for identifying an agent's actual beliefs.

This observation may at first seem to contradict the widely-accepted claim that the standard elicitation mechanisms are incentive-compatible. However, there is no contradiction really. What is implicitly assumed in most of the belief elicitation literature is that utilities are state-independent. And of course, in this case as we have already explained, the identification problem is assumed away.

\begin{example}\textsc{(Belief elicitation in the wife's problem).}\label{EX:belief elicitation mechanism}
Recall the wife's problem, and consider one of the most common elicitation mechanisms, viz., a quadratic scoring rule. Accordingly, the (risk-neutral) wife is asked to report a belief $r\in\Delta(S)$. In return she will receive a state-contingent payment of $\$100\bigl(1-(1-r(s_1))^2\bigr)$ in case her husband indeed recovers (i.e., at state $s_1$), and $\$100\bigl(1-(1-r(s_2))^2\bigr)$ in case he remains paralyzed (i.e., at state $s_2$). In other words, she will get a flat fee of \$100 minus a penalty which is proportional to the squared distance between her report and the realized state. Her preferences over reports are represented by the SEU
\begin{equation}\label{EQ:husband problem SEU}
\mathbb{E}_\mu\bigl(u(r)\bigr)=\mu(s_1)\gamma_1\Bigl(1-\bigl(1-r(s_1)\bigr)^2\Bigr)+\mu(s_2)\gamma_2\Bigl(1-\bigl(1-r(s_2)\bigr)^2\Bigr).
\end{equation}
But unfortunately we do not know the parameter specification $(\gamma_1,\gamma_2)$. As a result, the formula 
\begin{equation}
\frac{\mu(s_1)}{\mu(s_2)}=\frac{r(s_1)}{r(s_2)}\cdot\frac{\gamma_2}{\gamma_1},
\end{equation}
which we obtain from the first order condition, is completely uninformative to us. All that it says is that, upon hearing the report $r$, we first need to exogenously assume some values for the utilities parameters, and subsequently pin down the beliefs. But of course, if we use misspecified parameters, we will obtain wrong beliefs. For instance, suppose that we adopt the usual assumption that utilities are state-independent, i.e., assume $\gamma_1=\gamma_2$. Then, we will always take her actual beliefs to be the same as her reported ones. But unfortunately, in our example, consistently with her willingness to buy the insurance, the wife will actually misreport $r(s_1)=90\%$ (see Equation (\ref{EQ:Example husband})). In fact, the parameters that yield her actual beliefs of $\mu(s_1)=10\%$ must satisfy $\gamma_1=81\gamma_2$, rather than $\gamma_1=\gamma_2$. But again, this choice of parameters is completely arbitrary, meaning that the quadratic scoring rule is completely uninformative.
\end{example}
 
Then, we can go ahead and formally define the notion of the agent having no stakes in $S$. 
 
\begin{definition}\label{D:no stakes}\textsc{(No stakes).}
We say that the agent has no stakes in $S$ whenever any incentive-compatible mechanism elicits the her actual beliefs about $S$.
\end{definition}

Of course, the no-stakes condition is simply a renaming of the assumption of state-independent utilities, and in this sense it is non-testable with traditional choice data (due to the identification problem). Nevertheless, throughout the paper we would rather say that ``the agent has no stakes", instead of saying that ``the agent has state-independent utilities". This is because the former is expressed in terms of primitive ---albeit unobservable--- concepts, whereas the latter is expressed in terms of utilities, which are not even a primitive concept.

\section{Belief identification by proxy}\label{S:solution by proxy}

\subsection{Extending the state space}\label{S:extending the state space}

Having realized that the early solution to the identification problem (viz., to simply assume no stakes in the state realization) is unsatisfactory, decision theorists have taken an alternative general approach which relies on leveraging non-traditional choice data. There is a number of ways one could do that \citep[e.g.,][]{Dreze1961,Dreze1987, Fishburn1973, KarniSchmeidlerVind1983, Karni1992, Lu2019}, but unfortunately the data in most of these methods are complex and demanding. 

In this paper, we propose a new approach that will turn out to solve the identification problem in a theoretically sound, but at same time surprisingly straightforward way. The idea is to continue using traditional choice data, albeit over an appropriately extended state space. Formally, we introduce another state space $T=\{t_1,\dots,t_N\}$ with $|T|\geq |S|$, and we define the extended product space $S\times T$. The agent's actual belief over the extended state space is denoted by $\pi\in\Delta(S\times T)$.\footnote{The decision-theoretic foundations of this joint belief are presented in detail in Section \ref{S:foundations SEU extended}.}  For any nonempty $A\subseteq S$ and $E\subseteq T$, we respectively denote the marginal conditional beliefs $\pi_T(E|A):=\pi(A\times E|A\times T)$ and $\pi_S(A|E):=\pi(A\times E|S\times E)$. Marginal (unconditional) beliefs are simply denoted by $\pi_T$ and $\pi_S$ respectively.

\begin{definition}\textsc{(Proxy).}
We say that $T$ is a proxy for $S$ whenever the following are satisfied:
\begin{itemize}
\item[$(P_1)$] \textsc{Objective marginal beliefs:} The marginal belief $\pi_T$ is commonly known.
\item[$(P_2)$] \textsc{Uninformative event:} There is some event $E\subseteq T$ such that $\pi_S(\cdot|E)=\mu$.
\item[$(P_3)$] \textsc{Strong correlation:} For any two disjoint events $A,B\subseteq S$, we have $\pi_T(\cdot|A)\neq \pi_T(\cdot|B)$.
\end{itemize}
\end{definition}

In most examples throughout the paper, $(P_1)$ will typically reflect the idea that the distribution of $T$ is controlled by the experimenter, and has been publicly announced to the agent (Examples \ref{EX:experimental drug} and \ref{EX:expert charlatan I}). But there are also cases where $T$ describes the partition of a population with respect to some characteristic whose distribution is commonly known even without the experimenter announcing it (Example \ref{EX:pharmaceutical's problem}).

Condition $(P_2)$ states that introducing $T$ and conditioning with respect to $E\subseteq T$ does not provide any information about $S$ in comparison to the benchmark case where $T$ is not introduced in the first place. In this sense, $E$ is called the uninformative event. Notice that the event $E$ is not necessarily equal to $T$ itself, meaning that the actual belief $\mu$ does not necessarily coincide with the unconditional marginal $\pi_S$ (Examples \ref{EX:experimental drug} and \ref{EX:expert charlatan I}). Note that there is a conceptual similarity to the literature on growing awareness \citep[e.g.,][]{KarniViero2013}, in the sense that at first the agent may not even be aware of $T$, thus holding beliefs only about $S$, and only after we announce the proxy she holds beliefs about $S\times T$ (Examples \ref{EX:experimental drug} and \ref{EX:expert charlatan I}).

Finally, $(P_3)$ is perhaps the least-obvious of the three conditions. Loosely speaking, it guarantees that there is no redundant information within $T$, in the sense that $(P_3)$ is formally equivalent to the vectors $\pi_T(\cdot|s_1),\dots,\pi_T(\cdot|s_K)$ being linearly independent in $\mathbb{R}^T$ (see proof of Lemma \ref{L:main Theorem}). This characterization is very handy, as it allows us to directly test $(P_3)$. Finally, note that whenever $S$ is binary, this condition reduces to $\pi_T(\cdot|s_1)\neq\pi_T(\cdot|s_2)$, i.e. $S$ and $T$ are not independent. 

Let us illustrate the previous conditions with some examples related to the wife's problem. 

\begin{example}\label{EX:experimental drug}\textsc{(New drug).}
Recall the wife's problem from the previous section, and suppose that there is a promising new experimental drug in the market which is believed to expedite the recovery time from Guillain-Barr\'{e}. The husband is eligible to participate in a clinical trial. The proxy describes the two possible groups in which the husband can be placed:
$$T=\{\mbox{treatment group }(t_1), \mbox{ control group }(t_2)\}.$$
The wife knows that the chances of him being included in the treatment group are 50\%, and therefore her beliefs over $T$ are objective, in accordance with $(P_1)$. Moreover, she knows that in case he receives the placebo, his chances of recovery will remain unchanged, i.e., $\pi_S(\cdot|t_2)=\mu$. Here, we implicitly assume that the husband does not even know that he is participating in the clinical trial, and therefore his health cannot be affected by psychological factors, such as the well-known placebo effect. This means that, from the wife's point of view, $\{t_2\}$ is the uninformative event of $(P_2)$. Finally, based on all the previous studies, she has good reason to believe that the drug helps. Hence, the chances that he has taken the drug are higher if he recovers than if he does not, i.e., consistently with $(P_3)$, we have $\pi_T(\cdot|s_1)\neq \pi_T(\cdot|s_2)$. So overall, $T$ is a proxy for $S$.
\end{example}

\begin{example}\label{EX:expert charlatan I}\textsc{(Expert versus charlatan).}
Once again, we consider the wife's problem, supposing now that the wife is told that there are some good news, i.e., her husband's file was examined by some doctor who subsequently predicted that he will recover. The proxy describes the two possible expertise levels of the doctor that provided the good news:
$$T=\{\mbox{expert }(t_1), \mbox{ charlatan }(t_2)\}.$$
The wife is told that the probability of the doctor being an expert is 50\%, and therefore $(P_1)$ is satisfied. Moreover, once again $\{t_2\}$ is the uninformative event of $(P_2)$. This is because conditional on the diagnosis coming from a charlatan, the wife does not update her beliefs, i.e., $\pi_S(\cdot|t_2)=\mu$. Finally, the probability that the good news actually came from the expert is larger if the husband recovers than if he does not, i.e., $\pi_T(\cdot|s_1)\neq\pi_T(\cdot|s_2)$. Hence, $(P_3)$ is also satisfied. So once again, $T$ is a proxy for $S$.
\end{example}

\begin{example}\label{EX:pharmaceutical's problem}\textsc{(The CEO's problem).} 
Let us now look at the problem from the point of view of the CEO of the pharmaceutical company which is organizing the clinical trial for the new drug of Example \ref{EX:experimental drug}. We want to identify the CEO's expected number of patients that will recover among those that take the drug. Formally, this is equal to the probability that a randomly chosen patient recovers multiplied by the sample size. Obviously, the CEO has stakes in the drug being effective, and therefore the identification problem kicks in. Consider a proxy that partitions the subjects in two age groups:
$$T=\{\mbox{young }(t_1), \mbox{ old }(t_2)\}.$$
The CEO knows the age distribution within the treatment group. Hence, $(P_1)$ is satisfied. Then, it is straightforward to verify that $\mu=\pi_T(t_1)\pi_S(\cdot|t_1)+\pi_T(t_2)\pi_S(\cdot|t_2)$, meaning that $T$ is the uninformative event of $(P_2)$. Finally, the CEO expects age to be highly correlated with recovery probability, in the sense a higher proportion of young people is found among those who recover than among those who do not, and vice versa for old people, i.e., formally, $\pi_T(t_1|s_1)>\pi_T(t_1|s_2)$ and $\pi_T(t_2|s_1)<\pi_T(t_2|s_2)$. Thus, $(P_3)$ is also satisfied.
\end{example}

The previous three examples are representatives of the main families of natural proxies that we have in mind, viz., influential actions (Example \ref{EX:experimental drug}), evidence with stochastic reliability (Example \ref{EX:expert charlatan I}), representative sample (Example \ref{EX:pharmaceutical's problem}). Let us briefly elaborate on each of them:
\begin{itemize}
\item An \textsc{influential action} is an intervention that affects the realization of the state space \citep{Tsakas2020}. Conditional on an influential action having been taken, the agent's subjective beliefs are updated in a way unknown to us. On the other hand, conditional on not having taken any action, the agent's beliefs remain unchanged. Then, the idea is to stochastically take an influential action with probabilities controlled by the experimenter. 

\item \textsc{Evidence with stochastic reliability} consists of different experiments, one of which is uninformative. Then, we collect a piece of evidence and we communicate to the agent the probability of each of the experiments having generated this evidence. Moreover, conditional on the evidence coming from the informative experiment she would update her beliefs, whereas conditional on the evidence coming from the uninformative one she maintains her original beliefs. 

\item A \textsc{random sample} is used to identify the agent's expected belief about a certain variable in a population of individuals. The proxy partitions the population into equivalence classes with respect to some (demographic) characteristic. The proportion of each class in the population is assumed to be commonly known. The agent's conditional belief about the proportion of each class given each realization of the original variable can be identified, as the agent does not have stakes in the demographic.   
\end{itemize}

The existence of such abundance of potential proxies that one can pick directly off the shelf, provides us with great flexibility and confidence that we can almost always find one that can be practically implemented (see Section \ref{S:finding suitable proxies}). And it is exactly this flexibility that will make our approach appealing.


\subsection{Main identification result}\label{S:main identification result}

Let us now explain how we can leverage proxies to identify the agent's actual beliefs about the original state space $S$. First, we will need to make sure that the proxy is suitable. 

\begin{definition}\label{D:suitable proxy}\textsc{(Suitable proxy).}
A proxy $T$ is called suitable if the agent has no stakes in its realization conditional on $S$.
\end{definition}

\noindent In the context of our running examples, this is a natural assumption. For instance, conditional on the husband health condition, the wife will not care if he has taken the drug or the placebo (in Example \ref{EX:experimental drug}) or if the good diagnosis has come from the expert or the charlatan (in Example \ref{EX:expert charlatan I}). Likewise, conditional on a randomly chosen subject condition, the CEO does not care about how old this person is (Example \ref{EX:pharmaceutical's problem}). Of course, ex ante the agent may still care about these proxies, but only due to their correlation with the state space. But once uncertainty about state space has been resolved, she has no longer any stakes or interest in the realization of the proxy.

The fact that the agent does not have stakes in $T$ conditional on $S$ means that any incentive-compatible mechanism will elicit the agent's actual beliefs $\pi_T(\cdot|s)$ for all $s\in S$. And as it turns out, this suffices for (indirectly) identifying the agent's beliefs about $S$, which is what we wanted to achieve in the first place.\footnote{Our approach bears some similarity to the one that \cite{Thaler2020} uses in order to identify motivated reasoning: he also elicits the agent's subjective probability of a news source being truthful in order to test whether some piece of information about $S$ is internalized differentiably by people of different political affiliations. Still, his method is fundamentally different from ours, not only in the terms of the underlying question, but also in that he needs to directly elicit the (median) belief about $S$, which based on our entire discussion will likely suffer from the identification problem. Furthermore, his mechanism applies only to stochastic evidence, like our proxy in Example \ref{EX:expert charlatan I}.} The crucial intermediate step towards identification is to first pin down the joint belief $\pi$.

\begin{lemma}\label{L:main Theorem}
Let $T$ be a proxy for $S$ such that the agent does not have stakes in $T$ conditional on $S$. Then, her actual joint belief $\pi\in\Delta(S\times T)$ is identified by 
\begin{equation}\label{EQ:joint probability pi}
\pi(s,t)=\pi_S(s)\pi_T(t|s), 
\end{equation}
where $\pi_T(\cdot|s)$ is directly elicited for each $s\in S$ with any standard  incentive-compatible mechanism, and $\pi_S$ is the unique solution to
\begin{equation}\label{EQ:law of total probability}
\pi_T=\sum_{s\in S}\pi_T(\cdot|s)\pi_S(s).
\end{equation}
\end{lemma}

\begin{theorem}[Main identification result]\label{T:Main Theorem}
Let $T$ be a proxy for $S$ such that the agent does not have stakes in $T$ conditional on $S$. Then, the actual belief $\mu\in\Delta(S)$ is identified by
\begin{equation}\label{EQ:identified beliefs general formula}
\mu(s)=\frac{\pi(\{s\}\times E)}{\pi_T(E)},
\end{equation} 
where $E\subseteq T$ is the uninformative event, and $\pi\in\Delta(S\times T)$ is identified in Lemma \ref{L:main Theorem}.
\end{theorem}

In the context of Example \ref{EX:experimental drug}, suppose that the wife reports probability $\pi_T(t_2|s_1)=0.125$ to the husband having taken the placebo given that he recovers, and $\pi_T(t_2|s_2)=0.75$ to having taken the placebo given that he remains paralyzed. So, by solving (\ref{EQ:law of total probability}), we obtain $\pi_S(s_1)=0.40$. Hence, by (\ref{EQ:joint probability pi}), the joint belief $\pi\in\Delta(S\times T)$ is given by the following table.
\begin{center}
\begin{tikzpicture}[scale=1.9]
\draw[line width=0.8pt] (0,1) -- (0,2) -- (2.5,2) -- (2.5,1) -- (0,1);
\draw (0,1.3) node[left] {\footnotesize{placebo $(t_2)$}};
\draw (0,1.7) node[left] {\footnotesize{drug $(t_1)$}};
\draw (0,2) node[above right] {\footnotesize{recovers $(s_1)$}};
\draw (1.25,2) node[above right] {\footnotesize{paralyzed $(s_2)$}};
\draw (0.6,1.7) node {\footnotesize{$0.35$}};
\draw (1.9,1.7) node {\footnotesize{$0.15$}};
\draw (0.6,1.3) node {\footnotesize{$0.05$}};
\draw (1.9,1.3) node {\footnotesize{$0.45$}};
\end{tikzpicture}
\end{center}
Then, we can easily identify the wife's actual belief via $\mu(s_1)=\pi_S(s_1|t_2)=0.10$.

Note that Lemma \ref{L:main Theorem} allows us to identify the agent's conditional beliefs about $S$ given any non-empty subset of $T$. As a result, we can also identify how the agent incorporates a certain piece of information. For instance, in Example \ref{EX:experimental drug}, we can also identify how the wife updates her beliefs upon learning that her husband has taken the drug, i.e., $\pi_S(s_1|t_1)=0.70$. 

It is also important to emphasize that in order for our identification strategy to work, the cardinality of the proxy $T$ must be at least as large as the cardinality of the state space $S$. Otherwise, $(P_3)$ will be violated, since $\pi_T(\cdot|s_1),\dots,\pi_T(\cdot|s_K)$ will not be linearly independent, and a fortiori we will not be able to identify $\pi_S$ from Equation (\ref{EQ:law of total probability}). The good news is that the amount of data that we need in order to identify the agent's actual beliefs is quite small, especially in comparison with other existing methods, e.g., whenever the state space is binary, we only need to elicit two probabilities in order to identify the actual beliefs.

\subsection{Definition of subjective probability}\label{S:Definition of subjective probability}

As we have already mentioned early in the paper, the question of ``identifying the actual beliefs" is one that has been long debated. According to an early view within axiomatic decision theory, we should not even talk about actual beliefs, as the only observable primitive is the preference relation \citep[e.g.,][and references therein]{Karni2014}. In fact, beliefs only acquire meaning within a SEU model that disentangles them from utilities. Although we do not personally subscribe to this view,\footnote{As we have explained, this early approach is not satisfactory in many applications where we must treat actual beliefs as a primitive, and especially whenever we want to use this belief for purposes other than simply predicting the agent's choices among acts.} it can be easily reconciled with our main identification result. The idea is that we can interpret the belief $\mu$ that Theorem \ref{T:Main Theorem} identifies, simply as a parameter within an extended SEU model. And in this sense, $\mu$ could also alternatively serve as an alternative well-founded definition of subjective probability, without carrying the label of the ``actual belief".

\subsection{Identifying ``actual" utility functions}\label{S:identifying actual utilities}

Our earlier identification result allows us to obtain a well-founded definition of an ``actual" (state-dependent) utility function. This is a very useful object, as it allows to to give meaning to statements like ``state $s_1$ is preferred to state $s_2$" or the ``the utility of state $s$ is equal $w(s)$" (Section \ref{S:ranking the states}), which are in turn commonly used in many applied settings (e.g., in the literature on motivated beliefs).

Let us first illustrate the main idea of an ``actual" utility function in the context of the wife's problem. Recall, from Equation (\ref{EQ:Example husband}), that every SEU representation $(\tilde{u},\tilde{\mu})$ of $\succeq$ satisfies 
$$\frac{\tilde{\mu}(s_2)}{\tilde{\mu}(s_1)}=\frac{\gamma_1}{9\gamma_2},$$
where $\tilde{u}_1(q)=\gamma_1 q$ and $\tilde{u}_2(q)=\gamma_2 q$. This means that there is a duality between identification of beliefs and identification of relative marginal utilities. This exact duality is what the literature traditionally leverages to go from state-independent utilities to a unique belief, viz., if we set $\gamma_1=\gamma_2$ then we can uniquely identify $\bar{\mu}(s_1)=90\%$. In our case, on the other hand, we will use this duality the other way around, viz., using Theorem \ref{T:Main Theorem}, we have already identified the actual belief $\mu(s_1)=10\%$, and subsequently we can conclude that the utility parameters are such that $\gamma_1=81\gamma_2$. Hence, we can say that the ``actual" utility function is of the form
\begin{equation}\label{EQ:actual utilities}
u_1(q)=\alpha_1+81\beta q \mbox{ and }u_2(q)=\alpha_2+\beta q, 
\end{equation}
where $\alpha_1,\alpha_2\in\mathbb{R}$ and $\beta>0$. The following theorem generalizes this result beyond our simple example, even in cases where there is no state-independent representation.

\begin{theorem}\label{THM:actual utility function}
Let $(\tilde{u},\tilde{\mu})$ be an arbitrary SEU representation of the agent's preferences $\succeq$ over the set of acts $\mathcal{F}_S$. Moreover, the agent's actual belief $\mu\in\Delta(S)$ has been identified using Theorem \ref{T:Main Theorem}. Then, 
\begin{equation}\label{EQ:identification of actual utility function}
u_s=\alpha_s+\beta\frac{\tilde{\mu}(s)}{\mu(s)}\tilde{u}_s,
\end{equation}
with $\alpha_s\in\mathbb{R}$ and $\beta>0$, is the unique class of ``actual" utility functions such that the tuple $(u,\mu)$ is a SEU representation of $\succeq$. Moreover, this set of ``actual" utility functions remains invariant regardless of the initial SEU representation $(\tilde{u},\tilde{\mu})$ that we begin with.
\end{theorem}

It is important to point out that the ``actual" utility function remains invariant when the agent updates her beliefs upon observing new information about $S$. For instance, in Example \ref{EX:experimental drug}, suppose that the wife learns at some point that the husband has been indeed placed in the treatment group. Then, as shown in Section \ref{S:main identification result} (on the table below the main theorem), she will update her probability of him recovering to 70\%. Our Theorem \ref{T:Main Theorem} can still be used to identify this updated belief. Yet, the ``actual" utility functions that we obtain from Theorem \ref{THM:actual utility function} after the belief has been updated is still given by (\ref{EQ:actual utilities}). Of course, this is not surprising: in our Bayesian framework, the arrival of new information affects the beliefs, but not the tastes of the agent.

\section{Applications}\label{S:applications}

\subsection{Ranking the states}\label{S:ranking the states}


Using our identification results, we are now going to decompose the ``actual" utility into a part that represents the agent's (state-independent) tastes over consequences and and a part that represents her preferences over the state space. This way, for the second part, we will obtain a (cardinal) utility function over the state space.

Let us first assume that there exists a state-independent SEU representation, $(\bar{u},\bar{\mu})$. Of course, this does not mean that $\bar{\mu}$ coincides with the actual belief $\mu$. Still, it allows us to define the state-parameter 
\begin{equation}
w(s):=\frac{\bar{\mu}(s)}{\mu(s)},
\end{equation}
which will be henceforth interpreted as the cardinal utility of state $s$. Before explaining why we take $w(s)$ to be the utility for state $s$, note that by Theorem \ref{THM:actual utility function}, the agent's SEU can be rewritten as 
\begin{equation}\label{EQ:decomposition of SEU}
\mathbb{E}_\mu\bigl(u(f)\bigr)=\sum_{s\in S} \mu(s) \underbrace{w(s) \bar{u}(f_s)}_{u_s(f_s)}.
\end{equation}
In other words, the ``actual" state-utility can be decomposed into the utility $w(s)$ the agent obtains from state $s$ being realized, and the (state-independent) utility $\bar{u}(f_s)$ she obtains from the consequence she receives at $s$. 

Let us now go back to the interpretation of $w(s)$.  For simplicity and without loss of generality, assume that the set of consequences is a convex set of monetary payoffs $Q\subseteq\mathbb{R}$, and the utility function $u$ is statewise differentiable. Then, it is straightforward to verify that for any $q\in Q$ and any states $s,s'\in S$,
\begin{equation}\label{EQ:ranking of states equivalence with marginal utility}
w(s)\geq w(s') \ \Leftrightarrow \ u_s'(q)\geq u_{s'}'(q).
\end{equation} 
This follows directly from the decomposition in (\ref{EQ:decomposition of SEU}), which implies that $u_s'(q)=w(s)\bar{u}'(q)$. As a result, $w$ represents the relation between marginal utilities across states (using the ``actual" utility function). Actually, this interpretation of preferences over states is consistent with the one which is heuristically used throughout the literature. Thus our definition of utilities over states is quite natural. For instance, in our running example, as we established in Equation (\ref{EQ:actual utilities}), the wife's ``actual" marginal utility is larger at $s_1$ than at $s_2$, i.e., she prefers an extra dollar in a world where her husband has recovered to an extra dollar in a world where he is paralyzed. In this sense, we will say that she prefers state $s_1$ over state $s_2$, and this is represented by $w(s_1)>w(s_2)$. 

Interestingly, in this case where a state-independent SEU representation exists, the utility function $w$ is cardinal, meaning that we can also do interesting comparative statics using the decomposed SEU in (\ref{EQ:decomposition of SEU}).

It is not difficult to verify that the agent will be indifferent across all states if and only if the ``actual" utility function is state-independent. Not surprisingly, this equivalence is also consistent with our earlier notion of ``no stakes in the state realization" (Definition \ref{D:no stakes}).

Now, let us assume that there is no state-independent SEU representation. Then, we will not be able to decompose the ``actual" utility function into preferences over states and state-independent tastes, \`{a} la (\ref{EQ:decomposition of SEU}). Thus, it is impossible to obtain a cardinal utility function over the state space. Under certain conditions, we may still be able to obtain an ordinal utility function, i.e., (\ref{EQ:ranking of states equivalence with marginal utility}) may still hold. But even this is not certain, e.g., there can exist $p,q\in Q$ such that $u_s'(p)> u_{s'}'(p)$ and $u_s'(q)< u_{s'}'(q)$, meaning that state $s$ is preferred to $s'$ at $p$, but vice versa at $q$.

Finally, we should emphasize that ranking the states is not something unique to our paper; it can be done in principle in every model that uniquely pins down a unique state-dependent SEU representation. However, this is rarely done in practice. The reason is that there is a dichotomy in the literature: whenever a state-independent SEU representation exists, it is automatically assumed that the agent is indifferent across states; on the other hand, whenever there is no state-independent SEU representation, the ranking of the states may depend on the consequence at which states are evaluated. In our paper, there is no such dichotomy, because we recognize that the actual beliefs can be different from the beliefs given from the state-independent representation (even when the latter exists), and therefore the utility over states $w$ is not trivially constant.

\subsection{Identifying motivated beliefs}\label{S:motivated beliefs}

It is widely accepted in the literature that people hold \textit{motivated beliefs}, i.e., they systematically report biased beliefs in the direction of the state they prefer \citep{Kunda1990, Benabou2015}. For instance, people reportedly overestimate their own skill or performance \citep{Svenson1981, Zimmermann2020} or the chances of political events that they like \citep{Bullock2015, Thaler2020, Doiron2022}, even when they are incentivized to report truthfully. There are many explanations for this phenomenon in the literature, mostly revolving around the idea that (reported) beliefs are somehow self-serving. However, one should be careful with interpreting these results, because there is a fundamental measurement problem.

The issue is that motivated beliefs are typically encountered in environments where the agents have stakes in the state realization. However, as we have repeatedly explained throughout the paper, these are exactly the same environments where the belief identification problem arises. So, if we use traditional elicitation mechanisms, belief distortions due to motivated reasoning will be confounded with measurement errors due to falsely assuming state-independent utilities. And this gives rise to the following simple question: do people actually hold self-serving beliefs, or they simply misreport as a rational response to the incentives our elicitation task provides to them, or even perhaps something in between is happening? Let us illustrate this point in the wife's example.

\begin{example}\textsc{(Wife holding motivated beliefs).} 
Suppose that before we try to identify the wife's belief about her husband's recovery, we show all the information that she has ever received about Guillain-Barr\'{e} and her husband's health to some world-renowned expert. Assume that this expert estimates the probability of the husband recovering to 10\%. Subsequently, we elicit the wife belief using a standard mechanism. As we have already discussed, the wife will report a belief of 90\% (Example \ref{EX:belief elicitation mechanism}). Then, in most cases, the experimental economist would rush to conclude that the wife holds motivated beliefs, arguing that she did not take into account evidence that pointed in the direction of her husband remaining paralyzed, and this is what caused the huge discrepancy between her belief and the expert's estimate. However, said discrepancy could also be an artefact caused by to the identification problem, in which case our conclusion of her being a motivated reasoner would have been false.
\end{example}

Fortunately, this is where our main result comes in handy, as it allows us to disentangle distortions that come from motivated reasoning from the aforementioned measurement error. The idea is that by identifying beliefs with our indirect method of Theorem \ref{T:Main Theorem}, instead of the usual direct elicitation approach, we can confidently attribute any belief distortions that we find to the usual explanations from the motivated belief literature, without being concerned about confounding issues.

\section{Belief identification in practice}\label{S:implementation}

As we have already emphasized, the appeal of our method comes from its simplicity and the flexibility that it offers. Even so, research has shown that implementation of elicitation tasks often presents obstacles. Here we would like to point out potential such issues, and explain how they can be addressed and mitigated in our context. We should already stress at the outset that our goal is not to provide a detailed experimental design, but rather to theoretically address some issues that will likely pop up whenever our method is used in practice.

\subsection{Finding suitable proxies}\label{S:finding suitable proxies}

A major question is whether it is practically feasible to find suitable proxies. Admittedly, this may sometimes seem difficult when it comes to influential actions, but on the other hand evidence with stochastic reliability is almost always reasonably easy to implement. Furthermore, random sampling is another plausible option in several cases. Let us illustrate this point with a couple of additional examples.

\begin{example}\textsc{(Beliefs about electoral outcomes).}\label{EX:electoral outcome}
A presidential candidate is interested in his advisor's beliefs about winning the upcoming elections. Obviously, the advisor has stakes in the election outcome, e.g., due to intrinsic preferences or even due to her own political aspirations. Therefore, the identification problem will quite likely kick in. 

\vspace{0.2\baselineskip} \noindent \textsc{Influential action:} One influential action that could potentially act as a suitable proxy in this case is the stochastic launch of an extra online campaign. In particular, suppose that the advisor is told that with some probability, the candidate will hire a social media company to persuade voters in some swing states. Then, the advisor is asked to state his conditional beliefs that the campaign has been actually implemented conditional on the candidate having won and conditional on the candidate having lost the election. Here, it is natural to assume that, conditional on the election outcome, the advisor does not care whether the online campaign has been implemented or not. This is relevant to her only to the extent that it influences the outcome. 

\vspace{0.2\baselineskip} \noindent \textsc{Evidence with stochastic reliability:} While the influential action above would theoretically do the job in solving the identification problem, it is probably quite costly to implement. So, one alternative would be to use evidence with stochastic reliability. In particular, suppose that the advisor is told that an analyst predicted that her candidate will win at least two of the major swing states. However, at the same time she is also told that the analyst is in fact a reliable political expert with probability 70\% and he is a computer bot that spits out random predictions with the remaining 30\% probability. Note that this setting is identical to the one in Example \ref{EX:expert charlatan I}.

\vspace{0.2\baselineskip} \noindent \textsc{Random sampling:} Now suppose that instead of wanting to identify the advisor's subjective probability about winning the election, the candidate wants to know the advisor's expectation of the percentage of votes that he will receive in some swing state. This is the same as the probability that a randomly drawn voter casts a ballot for our candidate. It is known by everyone within the campaign that 55\% of the electorate in this state consists of women. Incidentally, it is believed that, in comparison to men, women are more likely to support our candidate. This means that one can use gender as a proxy, and elicit the advisor's conditional beliefs of a random woman voting our candidate given that he has won the state, as well as given that he has lost the state. This setting is identical to the one in Example \ref{EX:pharmaceutical's problem}.
\end{example}

\begin{example}\textsc{(Beliefs about relative performance).}
Suppose that an economist wants to identify the subjective probability that an experimental subject assigns to winning a contest during a lab experiment. It is quite natural to expect that the subject will have stakes in the outcome of the contest, e.g., due to self-image concerns, or perhaps due to the existence of a monetary prize that is associated with winning the contest. So, the identification problem will likely arise. 

\vspace{0.2\baselineskip} \noindent \textsc{Influential action:} Let us first assume that the contest involves only neutral tasks such that the subject will only care about winning the prize. In this case, consider an influential action along the following lines: the subject is informed that with probability 50\% her opponents are trained in the underlying task before the contest. Thus, once the subject knows whether she has won or not, she should not care about her opponents having taken the extra training or not. 

\vspace{0.2\baselineskip} \noindent \textsc{Evidence with stochastic reliability:} Of course, the previous proxy will not work if the subject has image concerns, e.g., if the subject is a mathematics student and the task is a math problem. In this case, she will likely have a preference to have faced a better trained opponent conditional on each outcome of the contest. Nevertheless, we can instead again use evidence with stochastic reliability. In particular, suppose that we tell the subject that a small sample of opponents performed on average worse than herself, which should lead her to update her own winning probability upwards. However, we also tell her that this piece of evidence is only true with probability 50\%. Notice that, conditional on the outcome of the contest (i.e., conditional on her performance against the whole population of opponents), she will not care if a small sample of her opponents performed better or worse than her. Therefore, the conditions of our theorem will be satisfied.
\end{example}

\begin{example}\textsc{(Beliefs about being caught cheating).}\label{EX:free riding}
A train company wants to know the beliefs of free-riders that they will get inspected. It is obvious that free-riders have stakes in not being caught cheating. Hence, the identification problem would likely arise once again.

\vspace{0.2\baselineskip} \noindent \textsc{Influential action:} One influential action is to stochastically double the number of inspectors, which in turn will make it more likely that passengers are checked. Of course, once a free-rider knows whether she has been caught or not, she should not care if there have been more inspections or not. This is only relevant to her in relation to the possibility of being caught or not. Note the analogy between this setting and Example \ref{EX:experimental drug}.

\vspace{0.2\baselineskip} \noindent \textsc{Evidence with stochastic reliability:} Like in earlier examples, the aforementioned influential action works perfectly fine from a theoretical point of view. However, it might be cumbersome to implement in practice. Thus, we can potentially use evidence with stochastic reliability. For instance, we can inform the passenger that seventy three free-riders were caught the previous week, while at the same time letting her know that this piece of information is accurate with probability 50\%. It is not difficult to verify that once again this setting is identical to the one of Example \ref{EX:expert charlatan I}. 
\end{example}

\subsection{Framing evidence with stochastic reliability}

Although from the previous discussion it is pretty clear that evidence with stochastic reliability is almost always a suitable proxy, one should careful with how exactly to implement it. Let us explain the shuttle issue using Example \ref{EX:expert charlatan I}. The main question is how to implement the 50\% probability that the good diagnosis came from an expert. We illustrate the point by describing two potential designs.

\begin{design}\label{Design 1}
Assume that we first collect diagnoses of a number doctors, out of which half are experts and half are charlatans. The wife knows that \textit{the total proportion of experts} is 50\%. Moreover, she knows that the conditional probability of a charlatan giving good news is 50\% regardless of the true state. On the other hand, her subjective probability of an expert giving good news is $1-\varepsilon$ conditional on $s_1$, and $\delta$ conditional on $s_2$. Without loss of generality, assume that $\varepsilon=0.80$ and $\delta=0.20$ are both commonly known, e.g., data about the expert's reliability are publicly available and have been already given to the wife. Then, the wife's joint belief is given by the following table:

\begin{center}
\begin{tikzpicture}[scale=1.9]
\draw[line width=0.8pt] (0,1) -- (0,2) -- (2.5,2) -- (2.5,1) -- (0,1);
\draw (0,1.3) node[left] {\footnotesize{bad news $(\omega_2)$}};
\draw (0,1.7) node[left] {\footnotesize{good news $(\omega_1)$}};
\draw (0,2) node[above right] {\footnotesize{recovers $(s_1)$}};
\draw (1.25,2) node[above right] {\footnotesize{paralyzed $(s_2)$}};
\draw (0.6,1.7) node {\footnotesize{$0.40\mu(s_1)$}};
\draw (1.9,1.7) node {\footnotesize{$0.10\mu(s_2)$}};
\draw (0.6,1.3) node {\footnotesize{$0.10\mu(s_1)$}};
\draw (1.9,1.3) node {\footnotesize{$0.40\mu(s_2)$}};
\draw (1.25,0.9) node[below] {\footnotesize{expert $(t_1)$}};

\draw[line width=0.8pt] (4.5,1) -- (4.5,2) -- (7,2) -- (7,1) -- (4.5,1);
\draw (4.5,1.3) node[left] {\footnotesize{bad news $(\omega_2)$}};
\draw (4.5,1.7) node[left] {\footnotesize{good news $(\omega_1)$}};
\draw (4.5,2) node[above right] {\footnotesize{recovers $(s_1)$}};
\draw (5.75,2) node[above right] {\footnotesize{paralyzed $(s_2)$}};
\draw (5.1,1.7) node {\footnotesize{$0.25\mu(s_1)$}};
\draw (6.4,1.7) node {\footnotesize{$0.25\mu(s_2)$}};
\draw (5.1,1.3) node {\footnotesize{$0.25\mu(s_1)$}};
\draw (6.4,1.3) node {\footnotesize{$0.25\mu(s_2)$}};
\draw (5.75,0.9) node[below] {\footnotesize{charlatan $(t_2)$}};
\end{tikzpicture}
\end{center}

\noindent Now suppose that we randomly draw one of the doctors, and we reveal to the wife his diagnosis (without of course telling the wife whether he is an expert or a charlatan). For the sake of our example, let this diagnosis be good. As a result, the wife's prior belief about $T$ will be 
$$\pi_T(t_1|\omega_1)=\frac{0.40\mu(s_1)+0.10\mu(s_2)}{0.65\mu(s_1)+0.35\mu(s_2)},$$ 
which is obviously not known to us.\footnote{Here the term prior refers to the probability about $T$ before conditioning on $S$, which corresponds to the belief $\pi_T$ in Theorem \ref{T:Main Theorem}. It does not refer to the agent's beliefs before having heard the doctor's diagnosis.} Hence, $(P_1)$ is violated. Therefore, our identification theorem will not apply. Notably, this is the case even though we know the wife's beliefs about the expert's reliability (as captured by $\varepsilon$ and $\delta$). Obviously, if these two parameters were not known, we would know even less about her beliefs, and therefore $(P_1)$ would still be violated.
\end{design}

\begin{design}\label{Design 2}
Suppose now that we begin again with a group of doctors, but \textit{the wife does not know the overall distribution of experts and charlatans}. Instead, she will be eventually told that \textit{from all the good news that we received, half of it came from experts and half of it from charlatans}. Let us explain, why this difference is crucial. For starters, similarly to our first design, the wife knows that the conditional probability of a charlatan giving a good news is 50\% regardless of the true state. And once again, her subjective probability of an expert giving good news is $1-\varepsilon$ conditional on $s_1$, and $\delta$ conditional on $s_2$. Then, we draw one doctor among those that gave good news and we reveal to the wife his diagnosis. At this point, we inform her that 50\% of the good diagnoses that we received came from experts and the remaining 50\% came from charlatans. Of course she still does not know the total number of experts and charlatans that we sampled. As a result, her prior belief about $S$ is 
$$\pi_T(t_1|\omega_1)=0.50,$$
implying that $(P_1)$ is now satisfied. Of course, in this design we will not know the wife's belief about $T$ unconditional on $\Omega$. But this is irrelevant for our purposes, as our identification theorem will simply compare the prior $\pi_T(\cdot|\omega_1)$ with the posteriors $\pi_T(\cdot|\omega_1,s_1)$ and $\pi_T(\cdot|\omega_1,s_2)$.
\end{design}

From the previous discussion, it becomes obvious that in order to implement stochastic evidence, we need to use the second design. That is, whenever we say that ``the probability of the evidence being informative is 50\%", we have already conditioned on the specific piece of evidence. In other words, the probability of 50\% does not reflect the proportion of informative sources in general (Design \ref{Design 1}), but rather the proportion of informative sources among those that have given the realized piece of evidence (Design \ref{Design 2}).


\subsection{Belief updating}\label{S:calibrated beliefs}

A major issue that we have to confront in practice is that we rely on the assumption that the agent updates beliefs in a Bayesian manner. Although from a theoretical standpoint this is not difficult to defend, from an empirical point of view we would have to seriously address the possibility of updating biases if we were to actually use our mechanism for eliciting beliefs in an experimental setting.  

The general framework within which updating biases are typically studied in the literature was introduced by \cite{Grether1980}. This model embeds Bayesian updating as a special case, and it is flexible enough to capture all known updating biases \citep{Benjamin2019}. Formally speaking, it introduces two free parameters $c>0$ and $d>0$, which distort the true likelihoods and the prior beliefs respectively, i.e., likelihoods are raised to $c$ and priors are raised to $d$. Subsequently, beliefs are updated in the usual way, i.e., by taking the relative (distorted) likelihood of each state, weighted by the corresponding (distorted) priors. In our context, posterior beliefs would then be given by
\begin{equation}\label{EQ:Grether updating}
\pi_T(t|s)=\frac{\pi_T(t)^d \pi_S(s|t)^c}{\pi_T(t_1)^d \pi_S(s|t_1)^c+\dots+\pi_T(t_N)^d \pi_S(s|t_N)^c} .
\end{equation}

In general, $c$ measures the inference bias, i.e., to what extent the signal is taken into account: in this sense it is called the \textit{inference parameter}. Parameter $d$ measures the extent to which the prior is used in the formation of the posteriors, and it is therefore called \textit{base-rate parameter}. Obviously, whenever $c=d=1$, we are back to the Bayesian benchmark. Moreover, note that whenever the prior is uniformly distributed, the base-rate parameter drops out, and posterior beliefs are only potentially distorted by the inference parameter. Thus, it becomes significantly easier to identify $c$ (in the absence of $d$), which is why many updating experiments start with uniform prior beliefs. In our case, this is a very convenient observation, as the prior $\pi_T$ is controlled by the experimenter, and therefore can be set to be uniform like in most examples above.


Our approach to deal with the possibility of non-Bayesian updating relies on calibrating the updating parameters, $c$ and $d$ (the latter only in case $\pi_T$ is not uniformly distributed) using out-of-sample data, and subsequently identifying the agent's beliefs.\footnote{In practice, calibration of the updating parameter is typically done by comparing the Bayesian posterior with the one that the agent has reported, in settings where the prior and the likelihoods are exogenously set. The most common such task is the well-known urn experiment. For an overview of the extensive experimental literature, we refer once again to the recent review of \cite{Benjamin2019}.} Let us illustrate how this is done in the context of our running example (Figure \ref{FIG:proxies graphical representation}). By construction, we know the prior beliefs $\pi_T$. Moreover, the posteriors $\pi_T(\cdot|s_1)$ and $\pi_T(\cdot|s_2)$ have been directly elicited. Hence, via Equation (\ref{EQ:Grether updating}) above, for each $s_k$, we obtain
$$\frac{\pi_T(t_1|s_k)}{\pi_T(t_2|s_k)}=\biggl(\frac{\pi_T(t_1)}{\pi_T(t_2)}\biggr)^d \biggl(\frac{\pi_S(s_k|t_1)}{\pi_S(s_k|t_2)}\biggr)^c,$$
which can be subsequently rewritten as
$$\underbrace{\biggl(\frac{\pi_T(t_1|s_k)}{\pi_T(t_2|s_k)}\biggr)^{1/c} \biggl(\frac{\pi_T(t_2)}{\pi_T(t_1)}\biggr)^{d/c}}_{\delta_k}=\frac{\nu_k}{\mu_k},$$
where $\delta_k$ is known. Thus, solving the system with respect to $\mu$ and $\nu$ we obtain 
\begin{equation}
\mu_1=\frac{1-\delta_2}{\delta_1-\delta_2}.
\end{equation}
Notably, given that we have a closed-form solution for the beliefs, we can use the confidence intervals around the estimated parameters $c$ and $d$ to obtain confidence intervals around $\mu$. 

The overall idea is methodologically similar to the one used by \cite{Offerman2009}, who in a different context use out of sample data to calibrate parameters that allow them to subsequently identify beliefs. In particular, their aim is to identify beliefs (under usual state-independent utilities) using the standard quadratic scoring rule. Unfortunately, is not incentive-compatible due to non-linear risk-preferences and probability weighting. So, using out of sample data, they calibrate risk-preference and probability-weighting parameters, in order to debias the elicited beliefs.

	\appendix
	
	\numberwithin{equation}{section}
	
	\setcounter{lemma}{0}
	\setcounter{corollary}{0}
	\setcounter{theorem}{0}
	\setcounter{proposition}{0}
	\setcounter{rema}{0}
	\setcounter{defin}{0}
	
	\renewcommand{\thelemma}{\Alph{section}\arabic{lemma}}
	\renewcommand{\theproposition}{\Alph{section}\arabic{proposition}}
	\renewcommand{\thecorollary}{\Alph{section}\arabic{corollary}}
	\renewcommand{\thetheorem}{\Alph{section}\arabic{theorem}}
	\renewcommand{\therema}{\Alph{section}\arabic{rema}}
	\renewcommand{\thedefin}{\Alph{section}\arabic{defin}}

\section{Proofs}\label{S:proofs}

\begin{proof}[\textup{\textbf{Proof of Lemma \ref{L:main Theorem}}}]
Observe that $(P_3)$ holds if and only if $\pi_T(\cdot|s_1),\dots,\pi_T(\cdot|s_K)$ are linearly independent in $\mathbb{R}^S$. Indeed, for any two disjoint $A,B\subseteq S$, we have
\begin{equation}
\pi_T(\cdot|A)=\pi_T(\cdot|B) \Leftrightarrow\sum_{s\in S} \underbrace{\bigl(\pi_S(s|A)-\pi_S(s|B)\bigr) }_{a_s}\pi_T(\cdot|s)=0,
\end{equation}
where at least two coefficients $a_s$ are non-zero. Hence, there exists a unique $\lambda=(\lambda_1,\dots,\lambda_K)\in\mathbb{R}_+^K$ with $\lambda_1+\dots+\lambda_K=1$ such that 
\begin{equation}\label{EQ:proof of main Lemma 1}
\pi_T=\sum_{k=1}^K \lambda_k \pi_T(\cdot|s_k).
\end{equation}
Hence, $\pi_S$ is the unique $\lambda$ that satisfies (\ref{EQ:proof of main Lemma 1}). Recall that $\pi_T$ is known and $\pi_T(\cdot|s_1),\dots,\pi_T(\cdot|s_K)$ has been identified by a standard incentive-compatible elicitation task. Hence, we can directly compute $\pi_S$. Finally, we obtain $\pi(s,t)=\pi_S(s)\pi_T(t|s)$ by simply applying the chain rule of probability.
\end{proof}

\begin{proof}[\textup{\textbf{Proof of Theorem \ref{T:Main Theorem}}}]
By the definition of conditional probability we have 
\begin{equation}
\pi_S(s|E)=\frac{\pi(\{s\}\times E)}{\pi_T(E)}.
\end{equation}
And by $(P_2)$, we obtain $\mu=\pi_S(\cdot|E)$, which completes the proof.
\end{proof}

\begin{proof}[\textup{\textbf{Proof of Theorem \ref{THM:actual utility function}}}]
Throughout this proof, for notation simplicity, we index states by $k=1,\dots,K$. First, we will prove that, if $u$ is defined as in (\ref{EQ:identification of actual utility function}) then $(u,\mu)$ is a SEU representation. Take arbitrary $(\alpha_1,\dots,\alpha_K)\in\mathbb{R}^K$ such that $\sum_{k=1}^K\mu(s_k)\alpha_k=\alpha$, and $\beta>0$. Then, for all $f\in\mathcal{F}_S$,  
\begin{equation}
\mathbb{E}_\mu(u(f))=\alpha+\beta\mathbb{E}_{\tilde{\mu}}(\tilde{u}(f)),
\end{equation}
which completes this part of the proof.

\vspace{0.3\baselineskip} \noindent Let us now prove the converse, i.e., if $(\hat{u},\mu)$ is a SEU then $\hat{u}$ is necessarily defined as in (\ref{EQ:identification of actual utility function}). For each $k=1,\dots,K$, define the (convex) range of the original state-utility function, $Y_k:=u_k(Q)\subseteq\mathbb{R}$. Then, it is obvious that, the state-utility function $\hat{u}_k$ is obtained via a strictly increasing transformation of $u_k$, i.e., there is a continuous strictly increasing $\phi_k:Y_k\rightarrow \mathbb{R}$ such that 
\begin{equation}\label{EQ:proof Thm2 transformation}
\hat{u}_k=\phi_k\circ u_k. 
\end{equation}
This is because both $u_k$ and $\hat{u}_k$ represent the same state-preferences (see Section \ref{S:foundations state-dependent SEU}). 

We will now show that $\phi_k$ is linear. Take an arbitrary $y\in\inter(Y_1\times\cdots\times Y_K)$, and for each $k=1,\dots,K$ define $y^k\in\mathbb{R}^K$ by 
\begin{equation}\label{EQ:definition yk}
y_\ell^k=\begin{cases}
y_\ell & \mbox{if } \ell\neq k,\\
y_k+\nicefrac{\delta}{\mu(s_k)} & \mbox{if } \ell=k.
\end{cases}
\end{equation}
Since $y$ is an interior point, there exists some $\delta_y>0$ such that $y^k\in Y_1\times\cdots\times Y_K$ for all $\delta\in(-\delta_y,\delta_y)$ for every $k=1,\dots,K$. Hence, for each $k=1,\dots,K$, there exists an act $f^k\in\mathcal{F}_S$ such that, for every $\ell=1,\dots,K$, it is the case that
\begin{equation}\label{EQ:proof Thm2 state utilities}
u_\ell(f_\ell^k)=y_\ell^k, 
\end{equation}
Therefore, by construction, we obtain
\begin{equation}\label{EQ:proof Thm2 SEU 1}
\mathbb{E}_\mu(u(f^k))=\sum_{\ell=1}^K \mu(s_\ell)y_\ell^k=\sum_{\ell=1}^K \mu(s_\ell)y_\ell+\delta.
\end{equation}
Since the right hand-side does not depend on $k$, we have $f^1\sim\cdots\sim f^K$. Then, since $(\hat{u},\mu)$ is a SEU representation, we obtain
\begin{equation}\label{EQ:proof Thm 2 indifference}
\mathbb{E}_\mu(\hat{u}(f^1)=\cdots=\mathbb{E}_\mu(\hat{u}(f^K).
\end{equation}
By combining (\ref{EQ:proof Thm2 transformation}) and (\ref{EQ:proof Thm2 state utilities}), we get
\begin{equation}
\mathbb{E}_\mu(\hat{u}(f^k)=\sum_{\ell=1}^K \mu(s_\ell)\phi_\ell(y_\ell^k),
\end{equation}
and consequently, by (\ref{EQ:proof Thm 2 indifference}), it follows that for any two distinct $k,\ell=1,\dots,K$,
$$\sum_{m=1}^K\mu(s_m)\phi_k(y_m^k)=\sum_{m=1}^K\mu(s_m)\phi_\ell(y_m^\ell).$$
Using the definition of $y^k$ and $y^\ell$ from (\ref{EQ:definition yk}), the previous equation can be rewritten as
$$\mu(s_k)\phi_k(y_k+\nicefrac{\delta}{\mu(s_k)})+\mu(s_\ell)\phi_\ell(y_\ell)=\mu(s_k)\phi_k(y_k)+\mu(s_\ell)\phi_\ell(y_\ell+\nicefrac{\delta}{\mu(s_\ell)}).$$
Rearranging terms, dividing both sides with $\delta$, and taking right limits, yields
\begin{equation}\label{EQ:left derivatives}
\phi_{k+}'(y_k)=\lim_{\delta\uparrow0}\frac{\phi_k(y_k+\nicefrac{\delta}{\mu(s_k)})-\phi_k(y_k)}{\nicefrac{\delta}{\mu(s_k)}}=\lim_{\delta\uparrow0}\frac{\phi_\ell(y_\ell+\nicefrac{\delta}{\mu(s_\ell)})-\phi_\ell(y_\ell)}{\nicefrac{\delta}{\mu(s_\ell)}}=\phi_{\ell+}'(y_\ell).
\end{equation}
Note that the right derivatives are well-defined as the respective domains $Y_k$ and $Y_\ell$ are convex sets. We repeat this exercise with any $y'\in\inter(Y_1\times\cdots\times Y_K)$ which agrees with $y$ at all coordinates except $k$, i.e., $y_k\neq y_k'$ and $y_\ell= y_\ell'$ for all $\ell\neq k$. Thus, we obtain 
\begin{equation}\label{EQ:left derivatives1}
\phi_{k+}'(y_k')=\phi_{\ell+}'(y_\ell').
\end{equation}
But, since $y_\ell=y_\ell'$, it follows directly that, for any two $y_k,y_k'\in\inter(Y_k)$,
\begin{equation}\label{EQ:left derivatives2}
\phi_{k+}'(y_k)=\phi_{k+}'(y_k'),
\end{equation}
i.e., the right derivative of $\phi_k$ is constant in the interior of its domain. Therefore, together with continuity (including at the boundaries in case those belong to $Y_k$) it implies that $\phi_k$ is linear for all $k=1,\dots,K$. But then, by (\ref{EQ:left derivatives}), the slope of $\phi_k$ is the same for all $k=1,\dots,K$, which completes the proof.
\end{proof}

\section{Decision-theoretic foundations}\label{S:decision-theoretic foundations}

In this section, we will review the main axiomatizations of SEU (Sections \ref{S:foundations state-dependent SEU} and \ref{S:foundations state-independent SEU}), and subsequently we will formalize the axioms that we implicitly assume when we introduce a suitable proxy (Section \ref{S:foundations SEU extended}). 

Consider a grand state space $\Theta$ . For the purposes of this paper, it is without loss of generality to assume that $\Theta$ is finite. Moreover, take a convex set of consequence $Q\subseteq\mathbb{R}^L$. This setting accommodates a number of interesting special cases, viz., the consequences are the set of lotteries over a finite set of outcomes $X$, an interval of monetary payoffs, the feasible consumption bundles given a budget constraints, the different ways to split a pie among a number of individuals, etc.

The set of all acts is $\mathcal{F}:=Q^\Theta$. The constant act that induces the consequence $q\in Q$ at all states is simply denoted by $q\in\mathcal{F}$. Compound acts are defined in the usual way, i.e., $\lambda f+(1-\lambda)g$ induces the consequence $\lambda f_\theta+(1-\lambda)g_\theta$ at each state $\theta$, which by convexity of $Q$ is well-defined. For any two acts $f,g\in\mathcal{F}$ and any $A\subseteq\Theta$, take the act
$$(f_A g)(\theta)=\begin{cases}
f(\theta) & \mbox{if } \theta\in A,\\
g(\theta) & \mbox{if } \theta\notin A,
\end{cases}$$
i.e., $f_A g$ coincides with $f$ in $A$ and with $g$ everywhere else. Whenever we have a singleton $A=\{\theta\}$, we slightly abuse notation to simply write $f_\theta g:=f_{\{\theta\}} g$.

The agent has (weak) preferences $\succeq$ over $\mathcal{F}$. As usual $\succ$ and $\sim$ denote the asymmetric part (viz., strict preference) and the symmetric part (viz., indifference) respectively. We assume preferences to be non-trivial, i.e., there exist $f,g\in\mathcal{F}$ such that $f\succ g$. A state $\theta$ is called $\succeq$-null if for any three acts $f,g,h\in\mathcal{F}$ we have $f_\theta h \sim g_\theta h$. Throughout the paper, for all acts  $f,g,h\in\mathcal{F}$, we assume the preferences to satisfy the following axioms:
\begin{itemize}
\item[$(A_1)$] \textsc{Completeness:} $f\succeq g$ or $g\succeq f$
\item[$(A_2)$] \textsc{Transitivity:} if $f\succeq g$ and $g\succeq h$, then $f\succeq h$
\item[$(A_3)$] \textsc{Continuity:} $\{g\in\mathcal{F}: f\succeq g\}$ and $\{g\in\mathcal{F}: g\succeq f\}$ are both closed in $\mathcal{F}$
\end{itemize}
It is well known that these three axioms are necessary and sufficient for the existence of a continuous utility representation $v:\mathcal{F}\rightarrow\mathbb{R}$ \citep{Debreu1954}.

\subsection{State-dependent SEU representations}\label{S:foundations state-dependent SEU}

Assume that for all $f,g,h,h'\in\mathcal{F}$ and all $A\subseteq\Theta$ the preferences satisfy:
\begin{itemize}
\item[$(A_4)$] \textsc{Separability:} $f_A h\succeq g_A h$ if and only if $f_A h'\succeq g_A h'$
\end{itemize}

\begin{theorem}[\citealp{Wakker1989}]\label{THM: Wakker1989}
Preferences over acts satisfy $(A_1)-(A_4)$ if and only for every full-support belief $\tilde{\pi}\in\Delta(\Theta)$ there is a continuous state-dependent utility function $\tilde{u}:Q\rightarrow\mathbb{R}^\Theta$ such that
\begin{equation}\label{EQ:Wakker state-dependent SEU}
f\succeq g \ \Leftrightarrow \ \sum_{\theta\in\Theta} \tilde{\pi}(\theta) \tilde{u}_\theta(f_\theta) \geq \sum_{\theta\in\Theta} \tilde{\pi}(\theta)\tilde{u}_\theta(g_\theta).
\end{equation}
\end{theorem}

The previous theorem is based on a well-known intermediate result showing that the four axioms are necessary and sufficient for an additively separable utility representation \citep{Debreu1960}. More specifically, first we define the agent's conditional preferences $\succeq_A$ over $Q$ given a nonempty $A\subseteq\Theta$, by means of the following equivalence: $p\succeq_A q$ if and only if $p_A h\succeq q_A h$ for all $h\in \mathcal{F}$. Obviously, the separability axiom guarantees that conditional preferences are well-defined and satisfy $(A_1)-(A_3)$. Then, it can be shown that for every $\theta\in\Theta$, the state-preferences $\succeq_\theta$ have a state-utility representation $v_\theta:Q\rightarrow\mathbb{R}$, such that
\begin{equation}
v(f)=\sum_{\theta\in\Theta} v_\theta(f_\theta)
\end{equation}
represents $\succeq$. Therefore, we can take any full-support belief $\tilde{\pi}\in\Delta(\Theta)$ and rescale each state-utility function via $\tilde{u}_\theta:=v_\theta/\tilde{\pi}(\theta)$. As a result, we directly obtain $v(f)=\mathbb{E}_{\tilde{\pi}}(\tilde{u}(f))$, which completes the proof of the previous result. 

Now, let us strengthen separability by replacing it with the following axiom. In particular, for every $f,g,h\in\mathcal{F}$ and every $\lambda\in(0,1)$ we assume:
\begin{itemize}
\item[$(A_4')$] \textsc{Independence:} $f\succeq g$ if and only if $\lambda f+(1-\lambda)h\succeq \lambda f+(1-\lambda)h$
\end{itemize}

\begin{theorem}\label{THM: vNM}
Preferences over acts satisfy $(A_1)-(A_3),(A_4')$ if and only if for every full-support belief $\tilde{\pi}\in\Delta(\Theta)$ there is a linear state-dependent utility function $\tilde{u}:Q\rightarrow\mathbb{R}^\Theta$ such that
\begin{equation}
f\succeq g \ \Leftrightarrow \ \sum_{\theta\in\Theta} \tilde{\pi}(\theta) \tilde{u}_\theta(f_\theta) \geq \sum_{\theta\in\Theta} \tilde{\pi}(\theta)\tilde{u}_\theta(g_\theta).
\end{equation}
\end{theorem}

The previous result is particularly useful in cases where the set of consequences is the set of lotteries $\Delta(X)$ over a finite set of outcomes $X$. In this case, the conditional preferences $\succeq_\theta$ at each state $\theta\in\Theta$ have a vNM EU representation, and therefore ---by linearity--- the state utility $\tilde{u}_\theta$ is fully characterized by the Bernoulli utility function $\tilde{U}_\theta:X\rightarrow\mathbb{R}$.

\begin{corollary}[\citealp{KarniSchmeidlerVind1983}]\label{C:KarniSchmeidlerVind1983}
Whenever $Q=\Delta(X)$ for a finite set $X$, preferences over acts satisfy $(A_1)-(A_3),(A_4')$ if and only if for every full-suppport belief $\tilde{\pi}\in\Delta(\Theta)$ there is a state-dependent Bernoulli utility function $\tilde{U}:X\rightarrow\mathbb{R}^\Theta$ such that
\begin{equation}\label{EQ:vNM state-dependent SEU}
f\succeq g \ \Leftrightarrow \ \sum_{\theta\in\Theta} \tilde{\pi}(\theta)\sum_{x\in X} f_\theta(x) \tilde{U}_\theta(x)\geq \sum_{\theta\in\Theta} \tilde{\pi}(\theta)\sum_{x\in X} g_\theta(x) \tilde{U}_\theta(x).
\end{equation}
\end{corollary}

\begin{figure}[h!]
\centering
\subfigure[\textsc{Theorem \ref{THM: Wakker1989}:} The utility representation is additively separable but not necessarily linear.]{
\begin{tikzpicture}[scale=0.9]
\fill[opaque,white!90!black] (0,0) -- (4,0.33) .. controls (5,1) and (6,1.5) .. (7,4.5) -- (3,4.17) .. controls (2,1.16) and (1,0.7) ..  (0,0);
\draw[->,line width=0.8pt] (0,0) -- (5,2.5) node[above] {\small{$\theta_2$}};
\draw[->,line width=0.8pt] (0,0)  -- (6,-1.5) node[below] {\small{$\theta_1$}};
\draw[->,line width=0.8pt] (0,0) node[left] {\footnotesize{0}}  -- (0,6);
\draw[line width=0.8pt] (4,-1) node[below] {\footnotesize{1}} -- (7,0.5)  -- (3,1.5) node[above] {\footnotesize{1}};
\draw[line width=0.8pt] (0,0) -- (4,0.33) .. controls (5,1) and (6,1.5) .. (7,4.5) -- (3,4.17) .. controls (2,1.16) and (1,0.7) ..  (0,0);
\draw[line width=1pt] (0,0) -- (4,1.5) node[right] {\footnotesize{$\tilde{u}_1$}};
\draw[line width=1pt] (0,0) .. controls (1,1) and (2,2) .. (3,6) node[right] {\footnotesize{$\tilde{u}_2$}};
\filldraw[dashed] (3,-0.75) -- (5.7,0.6) circle(1.2pt) -- (2.7,1.35);
\draw (5.6,0.42) node[right] {\footnotesize{$f$}};
\filldraw[dashed] (3,-0.75) circle(1.2pt) -- (3,1.12) circle(1.2pt) -- (0,1.87)  node[left] {\footnotesize{$\tilde{u}_1(f_1)$}} circle(1.2pt);
\filldraw (3,0.25) circle(1.2pt);
\filldraw[dashed] (2.7,1.35) circle(1.2pt)  -- (2.7,4.9) circle(1.2pt) -- (0,3.55) circle(1.2pt);
\draw (0,3.65) node[left] {\footnotesize{$\tilde{u}_2(f_2)$}};
\filldraw (2.7,3.32) circle(1.2pt);
\filldraw[dashed] (5.7,0.6) -- (5.7,3.8) circle(1.2pt) -- (0,3.2) circle(1.2pt);
\draw (0,3.1) node[left] {\footnotesize{$v(f)$}};
\end{tikzpicture}
}
\quad \quad
\subfigure[\textsc{Theorem \ref{THM: vNM}:} The utility representation is linear, and a fortiori also additively separable.]{
\begin{tikzpicture}[scale=0.9]
\fill[opaque,white!90!black] (0,0) -- (4,0.33) -- (7,4.5) -- (3,4.17) -- (0,0);
\draw[->,line width=0.8pt] (0,0) -- (5,2.5) node[above] {\small{$\theta_2$}};
\draw[->,line width=0.8pt] (0,0)  -- (6,-1.5) node[below] {\small{$\theta_1$}};
\draw[->,line width=0.8pt] (0,0) node[left] {\footnotesize{0}} -- (0,6);
\filldraw[line width=0.8pt] (4,-1) node[below] {\footnotesize{1}} -- (7,0.5) node[right] {\footnotesize{$f$}} circle(1.2pt) -- (3,1.5) node[above left] {\footnotesize{1}};
\draw[line width=0.8pt] (0,0) -- (4,0.33) -- (7,4.5) -- (3,4.17) --  (0,0);
\draw[dashed] (7,0.5) -- (7,4.5) node[right] {\footnotesize{$v(f)$}};
\draw[line width=1pt] (0,0) -- (4,1.5);
\draw[line width=1pt] (0,0) -- (3,6);
\draw[dashed] (3,1.5) -- (3,6);
\draw[dashed] (4,-1) -- (4,1.5);
\filldraw (4,1.5) node[right] {\footnotesize{$\tilde{u}_1$}} circle (1.2pt);
\filldraw (3,6) node[right] {\footnotesize{$\tilde{u}_2$}} circle (1.2pt);
\draw[dashed] (3,1.5) -- (3,6) -- (0,4.5) node[left] {\footnotesize{$\tilde{U}_2(1)$}};
\draw[dashed] (4,-1) -- (4,1.5) -- (0,2.5) node[left] {\footnotesize{$\tilde{U}_1(1)$}};
\end{tikzpicture}
}
\caption{State-dependent SEU representation.}\label{FIG:State-dependent SEU}
\end{figure}
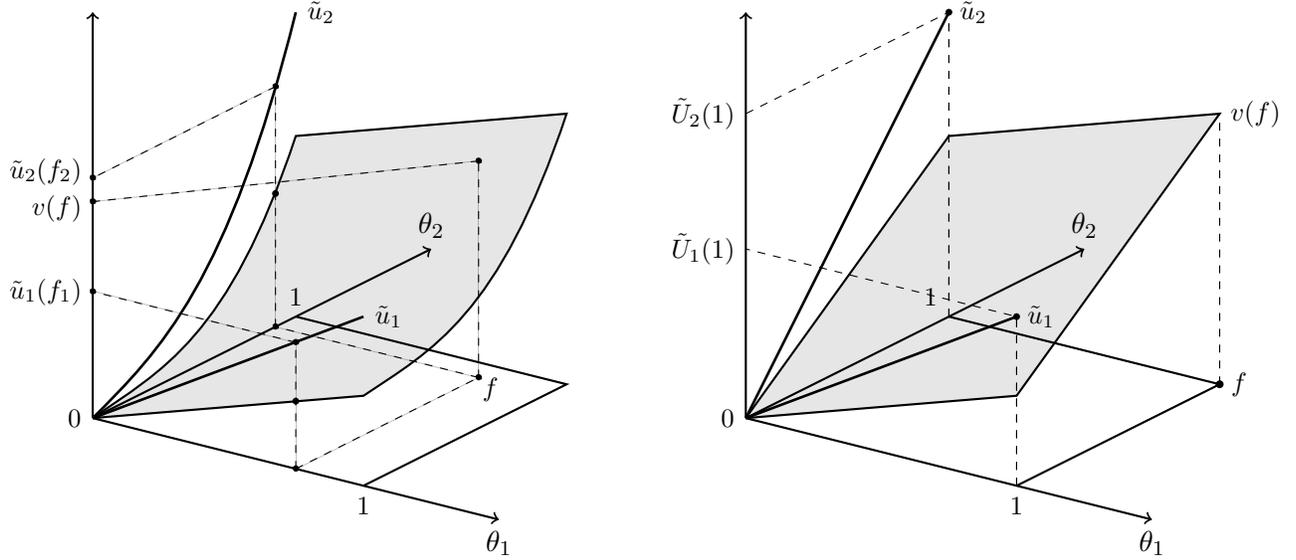

Let us illustrate the previous results graphically in a setting with a binary state space $\Theta=\{\theta_1,\theta_2\}$ and the set of consequences being the unit interval $X=[0,1]$ (Figure \ref{FIG:State-dependent SEU}).
On the left hand-side figure, the preferences satisfy only separability. Therefore, the state utility representation $v$ is not linear. The corresponding state-utilities are $v_1(q)=q$ and $v_2(q)=q^2$. On the right hand-side figure the preferences satisfy independence, and therefore $v$ is linear. Hence, it can be decomposed into linear state utilities, $v_1(q)=q$ and $v_2(q)=2q$. Since $v_k$ is linear, we can express $v_k(q)$ as a linear combination of the value of $v_k$ at the extreme points. 

What is interesting to point out is the duality between beliefs and state-utilities that we have already mentioned in Section \ref{S:identifying actual utilities}. The idea is that ---instead of first fixing the beliefs and then rescaling the state-utilities--- we can go the other way around, i.e., first take an arbitrary affine linear transformation $\tilde{u}_k:=\tilde{\beta}_k v_k$ of each state representation, and then pick the belief $\tilde{\pi}$ so that
\begin{equation}\label{EQ:from utilities to beliefs}
\tilde{\pi}(\theta_1):=\frac{\nicefrac{1}{\tilde{\beta}_1}}{\nicefrac{1}{\tilde{\beta}_1}+\nicefrac{1}{\tilde{\beta}_2}} \mbox{ and } \tilde{\pi}(\theta_2):=\frac{\nicefrac{1}{\tilde{\beta}_2}}{\nicefrac{1}{\tilde{\beta}_1}+\nicefrac{1}{\tilde{\beta}_2}}.
\end{equation}
Then, it is not difficult to verify that $v(f)=\bigl(\nicefrac{1}{\tilde{\beta}_1}+\nicefrac{1}{\tilde{\beta}_2}\bigr)\mathbb{E}_{\tilde{\pi}}(\tilde{u}(f))$ for any $f\in\mathcal{F}$, meaning that $(\tilde{u},\tilde{\pi})$ is a SEU representation. The general idea is that we can make a state-utility steeper and correspondingly decrease the relative probability of this state, without distorting the overall utility representation. 

Still it is important to notice that we can only use positive affine transformations of the state-utilities. The fact that transformations are affine means that the concavity of $\tilde{u}_\theta$ remains the same across different SEU  representations. Furthermore, the fact that these are positive transformations implies that the ordinal ranking of consequences remains the same for every state. Of course the latter is not surprising as $\tilde{u}_\theta$ will ---by construction--- represent $\succeq_\theta$.

\subsection{State-independent SEU representation}\label{S:foundations state-independent SEU}

Now, let us introduce additional axioms which will guarantee the existence of a state-independent representation. Let us start from the setting where we simply impose separability, but not necessarily independence. In this case, for any acts $f,g,h,h'\in\mathcal{F}$, any consequences $p,p',q,q'\in X$, and any non-$\succeq$-null states $\theta,\theta'\in\Theta$, we additionally assume that the preferences satisfy:
\begin{itemize}
\item[$(A_5)$] \textsc{State-independent preference intensity:} if $p_{\theta}f\succeq p'_{\theta}g$ and $q_{\theta}g\succeq q'_{\theta}f$ and $p_{\theta'}h\succeq p'_{\theta'}h'$, then $q_{\theta'}h\succeq q'_{\theta'}h'$.
\end{itemize}
This axiom is originally introduced by \cite{Wakker1989}, who shows that it eventually boils down to not having contradictory tradeoffs. Although it is stated independently of the remaining axioms, let us provide some intuition on what it says in a context where preferences have a SEU representation $(\tilde{u},\tilde{\pi})$, i.e., suppose that $(A_5)$ is imposed together with $(A_1)-(A_4)$. In this case, it postulates that
\begin{equation*}
\tilde{u}_\theta(p)-\tilde{u}_\theta(p')\geq \tilde{u}_\theta(q')-\tilde{u}_\theta(q) \mbox{ implies } \tilde{u}_{\theta'}(p)-\tilde{u}_{\theta'}(p')\geq \tilde{u}_{\theta'}(q')-\tilde{u}_{\theta'}(q).
\end{equation*}

\begin{theorem}[\citealp{Wakker1989}]
Preferences over acts satisfy $(A_1)-(A_5)$ if and only there exists a unique belief $\bar{\pi}\in\Delta(\Theta)$ and a continuous state-independent utility function $\bar{u}:Q\rightarrow\mathbb{R}$ such that
\begin{equation}\label{EQ:Wakker state-independent SEU}
f\succeq g \ \Leftrightarrow \ \sum_{\theta\in\Theta} \bar{\pi}(\theta) \bar{u}(f_\theta) \geq \sum_{\theta\in\Theta} \bar{\pi}(\theta)\bar{u}(g_\theta).
\end{equation}
\end{theorem}

Intuitively, $(A_5)$ guarantees that the different state-utilities are positive affine transformations of one another. And therefore, it is always possible to find a positive affine transformation for each $v_\theta$ so that the resulting $\bar{u}_\theta$'s are all the same. Then, we can find the unique belief $\bar{\pi}$ via Equation (\ref{EQ:from utilities to beliefs}). Clearly, this is not possible in Figure \ref{FIG:State-dependent SEU}.a, where the state-utilities are simply positive ---but not affine--- transformations of one another. In other words, in that case the state-preferences are the same, but the state-utilities are not. This weaker condition will turn out to be sufficient in cases where we have already strengthened separability to independence. 

Formally, for every act $f\in\mathcal{F}$, all consequences $p,q\in Q$, and all non-$\succeq$-null states $\theta,\theta'\in\Theta$, assume that the preferences satisfy:
\begin{itemize}
\item[$(A_5')$] \textsc{State-monotonicity:} $p_{\theta}f\succeq q_{\theta}f$ if and only if $p_{\theta'}f\succeq q_{\theta'}f$
\end{itemize}

\begin{theorem}[\citealp{AnscombeAumann1963}]
Preferences over acts satisfy $(A_1)-(A_3),(A_4')-(A_5')$ if and only if there is a state-independent linear utility function $\bar{u}:Q\rightarrow\mathbb{R}$ and a unique belief $\bar{\pi}\in\Delta(\Theta)$ such that
\begin{equation}
f\succeq g \ \Leftrightarrow \ \sum_{\theta\in\Theta} \bar{\pi}(\theta) \bar{u}(f_\theta) \geq \sum_{\theta\in\Theta} \bar{\pi}(\theta)\bar{u}(g_\theta).
\end{equation}
Moreover, whenever the set of consequences is $Q=\Delta(X)$ for some finite set $X$, there exists a Bernoulli utility function $\bar{U}:X\rightarrow\mathbb{R}$ such that 
\begin{equation}
\bar{u}(f_\theta)=\sum_{x\in X} f_\theta(x)\bar{U}(x)
\end{equation}
for each state $\theta\in\Theta$.
\end{theorem}

The key reason why we can weaken $(A_5)$ to $(A_5')$ is because the linear structure that our utility function inherits from $(A_4')$ guarantees that all state-utility functions will anyway share the same convexity properties. Therefore, we only need to guarantee that the preference ranking of consequences will remain the same across states. Actually, to see why $(A_5')$ is not enough without $(A_4')$, observe that it is already satisfied in Figure \ref{FIG:State-dependent SEU}.a, and nevertheless there is no state-independent SEU representation.

\subsection{Decision-theoretic foundations for proxies}\label{S:foundations SEU extended}

Let us now interpret $S$ as random variables over $\Theta$. That is, our original state space is formally a surjective mapping $\Theta\mapsto S$, and each $s\in S$  identifies the states $\theta\in\Theta$ that are mapped on $s$. In this case, $\mathcal{F}_S\subseteq\mathcal{F}$ is simply the set of $S$-measurable acts, i.e., $\mathcal{F}_S$ contains exactly those acts $f\in\mathcal{F}$ such that,  $f(\theta)=f(\theta')$ if and only if $\theta$ and $\theta'$ have the same image on $S$. Moreover, $\succeq_s$ denotes the conditional preferences given the pre-image of $s$ in $\Theta$. In exactly the same way, we interpret $T$ as another random variable, and we define the $T$-measurable acts $\mathcal{F}_T$. Finally, it is quite obvious that the product space $S\times T$ is likewise interpreted as a random variable, while $\mathcal{F}_{S\times T}$ denotes the set of $(S\times T)$-measurable acts. Without loss of generality, let us assume that $\Theta=S\times T$, meaning that every event in $\Theta$ is $(S\times T)$-measurable, and a fortiori $\mathcal{F}=\mathcal{F}_{S\times T}$. 

With the interpretation of $S$ and $T$ as random variables, we can formalize the notion of a proxy within a decision-theoretic framework. This will allow us to clarify, which parts in the definition of a proxy can be tested with traditional choice data, and which parts can only be exogenously assumed. Let us start with the following axioms:\footnote{Here we follow the axiomatization of \cite{Wakker1989}. We could have instead worked with \cite{AnscombeAumann1963}, in case we had $Q=\Delta(X)$, in which case in everything that follows we would have replaced $(A_4)$ with $(A_4')$ and $(A_5)$ with $(A_5')$.}
\begin{itemize}
\item[$(C_1)$] The preferences $\succeq$ satisfy $(A_1)-(A_4)$ for all acts in $\mathcal{F}$,
\item[$(C_2)$] For all $s\in S$, the conditional preferences $\succeq_s$ satisfy $(A_5)$ for all acts in $\mathcal{F}_T$.
\end{itemize}

Condition $(C_1)$ guarantees that there exists a state-dependent SEU representation $(\tilde{u},\tilde{\pi})$ of preferences over $\mathcal{F}$.\footnote{If we drop our simplification that $\Theta=S\times T$, then $(C_1)$ will simply postulate $(A_1)-(A_4)$ for the acts in $\mathcal{F}_{S\times T}$, and the corresponding state-dependent SEU representation will be a tuple $(\tilde{u},\tilde{\pi}_{S\times T})$.}  Condition $(C_2)$ guarantees that there exist ``partially" state-independent SEU representations. In this sense, our axiomatic system is much weaker than the one traditionally assumed, which would have simply postulated $(A_1)-(A_5)$ over the whole $\mathcal{F}$, and therefore it would have guaranteed existence of a state-independent representation globally. At the same time,  $(C_1)-(C_2)$ is stronger than simply requiring a state-dependent SEU. The latter would not have given us enough structure to circumvent the identification problem.

Notably, this is how far we can go with axiomatizing proxies. The rest of the way, we need to impose exogenously non-testable assumptions. 

For starters, $(P_1)-(P_2)$ are simply introduced based on intuitive appeal, and cannot be tested given that they are simply statements about the actual beliefs. In this sense, these are simply conditions that help us select some models from the entire class of SEU models that are consistent with $(C_1)-(C_2)$. 

The previous analysis applies partially to our condition that the agent has no stakes in $T$ conditional on $S$. Namely, axiom $(C_2)$ is necessary for the agent not to have stakes the proxy $T$ conditional on $S$, but of course it is not sufficient. As a result, we can only reject our no-stakes condition based on choice data, but we cannot guarantee that it holds. In this sense, we still need to exogenously assume it, like we do with $(P_1)$ and $(P_2)$.

Finally, this brings us to $(P_3)$, which can be tested under our earlier assumption that the agent has no stakes in $T$ conditional on $S$. In this sense, we do not need to impose any additional exogenous assumption.

Now, let us go back to our argument throughout the paper about our approach being appealing due to its flexibility. To see what we mean more clearly, relax the simplifying assumption that $\Theta=S\times T$, and assume that $\Theta$ is a rich set that can potentially describe all the potential proxies for $S$ as different random variables. In this case, our postulates $(C_1)-(C_2)$ will change depending on which random variable $T$ we have selected. This means that our conditions are always imposed ex post, relative to the variable $T$. And given that $\Theta$ is rich, we can always choose $T$ is a way that makes it easier for us to justify the aforementioned exogenous conditions for this specific $T$. On the flip side, the traditional approach will keep assuming $(A_1)-(A_5)$ together with a no-stakes condition for the entire state space. Clearly the latter is very restricting, which is what makes it difficult to justify.


\begin{small}

\end{small}

\end{document}